\documentclass[prx,aps,amssymb,twocolumn,longbibliography, nofootinbib,superscriptaddress]{revtex4-1}

\usepackage{amsmath}
\usepackage{amssymb}
\usepackage{amsthm}
\usepackage{amsfonts}
\usepackage{dsfont}
\usepackage{listings}
\usepackage{xcolor}
\usepackage[normalem]{ulem}
\usepackage{enumerate}
\usepackage{latexsym}

\usepackage{psfrag}

\usepackage{graphicx}
\usepackage{subfigure}

\newcommand{\beq}{\begin{equation}}
\newcommand{\eneq}{\end{equation}}

\newcommand{\bra}[1]{\left\langle#1\right|}
\newcommand{\ket}[1]{\left|#1\right\rangle}

\newcommand{\rhored}{\rho_{\rm red}}
\newcommand{\prodal}[2]{\underset{#1}{\overset{#2}{\prod}}}
\newcommand{\sumal}[2]{\underset{#1}{\overset{#2}{\sum}}}

\newcommand{\xe}{\xi_{\textrm{ext}}}
\newcommand{\xl}{\xi_{\textrm{loc}}}

\newcommand{\Jm}{J^\ast}
\newcommand{\hm}{h^\ast}

\input{epsf}

\newcommand{\twopartdef}[4]
{
	\left\{
		\begin{array}{ll}
			#1 & \mbox{if } #2 \\
			#3 & \mbox{if } #4
		\end{array}
	\right.
}

\newcommand{\twopartdefoth}[3]
{
	\left\{
		\begin{array}{ll}
			#1 & \mbox{if } #2 \\
			#3 & \mbox{otherwise}
		\end{array}
	\right.
}

\newcommand{\mo}{\mathcal{O}}

\newcommand{\mE}{\mathcal{E}}

\newcommand{\nn}{\nonumber}

\usepackage[colorlinks=true]{hyperref}
\hypersetup{citecolor = blue}

\begin{document}

\title{Perturbative instability towards delocalization \\at phase transitions between MBL phases}
\author{Sanjay Moudgalya}
\affiliation{Department of Physics, Princeton University, Princeton, NJ 08544, USA}
\author{David A. Huse}
\affiliation{Department of Physics, Princeton University, Princeton, NJ 08544, USA}
\author{Vedika Khemani}
\affiliation{Department of Physics, Stanford University, Stanford, CA 94305, USA}

\date{\today}

\begin{abstract}

We examine the stability of marginally Anderson localized phase transitions between localized phases to the addition of many-body interactions, focusing in particular on the spin-glass to paramagnet transition in a disordered transverse field Ising model in one dimension. We find evidence for a \emph{perturbative} instability of localization at finite energy densities once interactions are added, \emph{i.e.} evidence for the relevance of interactions - in a renormalization group sense - to the non-interacting critical point governed by infinite randomness scaling. We introduce a novel diagnostic, the ``susceptibility of entanglement", which allows us to perturbatively probe the effect of adding interactions on the entanglement properties of eigenstates, and helps us elucidate the resonant processes that can cause thermalization. The susceptibility serves as a much more sensitive probe, and its divergence can detect the perturbative beginnings of an incipient instability even in regimes and system sizes for which  conventional diagnostics point towards localization. 
We expect this new measure to be of independent interest for analyzing the stability of localization in a variety of different settings. 

\end{abstract}

\maketitle
\section{Introduction}
\label{sec:intro}
Many-body localization (MBL) was born in investigations of the stability of Anderson localization -- the phenomenon that strong enough disorder exponentially localizes non-interacting wavefunctions -- to the addition of interactions~\cite{Nandkishore2015, AbaninRMP2019}.  This stability was demonstrated to all orders in perturbation theory, following early precursors~\cite{anderson1958absence, Fleishman, basko2006metal}. While the \emph{non}-perturbative stability of MBL remains an open question in various settings~\cite{de2017stability,Luitzsmallbath, de2016absence, potirniche2019exploration}, it has been proven that a stable MBL phase can exist in local, strongly-disordered one dimensional spin chains~\cite{imbrie2016many}. 
More generally, understanding the stability of phenomena to small changes, such as the introduction of interactions, is a central enterprise in theoretical physics.
In this work, we add to this important literature by examining the effect of interactions on 
marginally Anderson localized critical points between Anderson localized phases.  

While phases and phase transitions are traditionally studied in the framework of equilibrium statistical mechanics, recent work has shown that there is a rich notion of phase structure even \emph{within} the out-of-equilibrium MBL phase~\cite{huse2013localization, pekker2014hilbert}. Different MBL phases represent distinct types of novel \emph{dynamical} phenomena that may be completely invisible to, or forbidden by, equilibrium thermodynamics --- a paradigmatic example being the recently discovered Floquet MBL time-crystal phase~\cite{khemani2016phase, Else2016, von2016absolute}. Localized phases can be understood as \emph{eigenstate phases} characterized by distinct patterns of long-range order (LRO), both symmetry-breaking and topological, in \emph{individual} highly-excited MBL eigenstates~\cite{huse2013localization, pekker2014hilbert, bauer2013area, chandran2014many, bahri2015localization, parameswaran2017eigenstate}; a phase transition between different localized phases requires singular changes in the eigenspectrum properties.
Indeed, the passage from localization to thermalization is itself a dynamical phase transition involving a singular change in the entanglement properties of highly excited many-body eigenstates. While the nature of the MBL-to-thermal phase transition has been a subject of intense study~\cite{pal2010many, luitz2015many, kjall2014many, vosk2015theory, pvp, deroeck_rg, clarkbimodal,khemani2017critical, goremykina2019analytically, morningstar2019renormalization, Vidmar2019, Abanin2019, Panda_2020, Sierant2020}, transitions between different MBL phases have recently received considerably less attention and are the focus of this work.

While our conclusions are quite general, for specificity, the majority of our analysis will be presented for a disordered transverse-field Ising model (TFIM) in one dimension. This model exhibits a localized, symmetry-broken ``spin-glass" (SG) phase with LRO, and a localized paramagnetic (PM) phase with no order~\cite{fisher1994random, fisher1995critical, huse2013localization, pekker2014hilbert, kjall2014many}. In the absence of interactions, the phase transition between the SG and PM phases is governed by an infinite-randomness fixed point that is studied using the strong disorder renormalization group (SDRG)~\cite{fisher1995critical, pekker2014hilbert}. While all single-particle (SP) eigenstates are exponentially localized in either phase, the critical point (CP) is only \emph{marginally} localized. The SP eigenstates corresponding to SP energies $\mathcal{E} \rightarrow 0$ are only \emph{stretched} exponentially localized at the CP, and both the SP density of states and localization length diverge in this limit~\cite{fisher1995critical}.

Once weak interactions are added, localization remains stable deep in the PM and SG phases, on account of the usual arguments for the stability of Anderson localization with strong enough disorder. The stability of localization near the CP, however, requires careful consideration. The CP exhibits ``marginal" Anderson localization due to the presence of (weakly) extended states in the SP spectrum, which could aid in the formation of long-range ``resonances" that make the CP more susceptible to thermalization~\cite{nandkishore2014marginal, de2017stability}. 
We note that the instability towards thermalization may be visible in a perturbative treatment of the interactions~\cite{basko2006metal, nandkishore2014marginal}, or it may have a subtler non-perturbative origin, in which case it will only be detectable at asymptotically large system sizes and times~\cite{de2017stability}. Previous SDRG studies of the interacting Ising model explicitly treat interactions as irrelevant -- even at finite many-body energy densities --  and so do not consider the resonances that could destabilize the CP~\cite{vosk2014dynamical, pekker2014hilbert}. The stability of this transition \emph{i.e.}  the relevance of interactions to the non-interacting CP in RG language, has been a major outstanding question in the literature and is the subject of this work. We find evidence in favor of a \emph{perturbative} instability of the CP to the addition to arbitrarily weak interactions.

\begin{figure}[t]
\includegraphics[scale = 0.9]{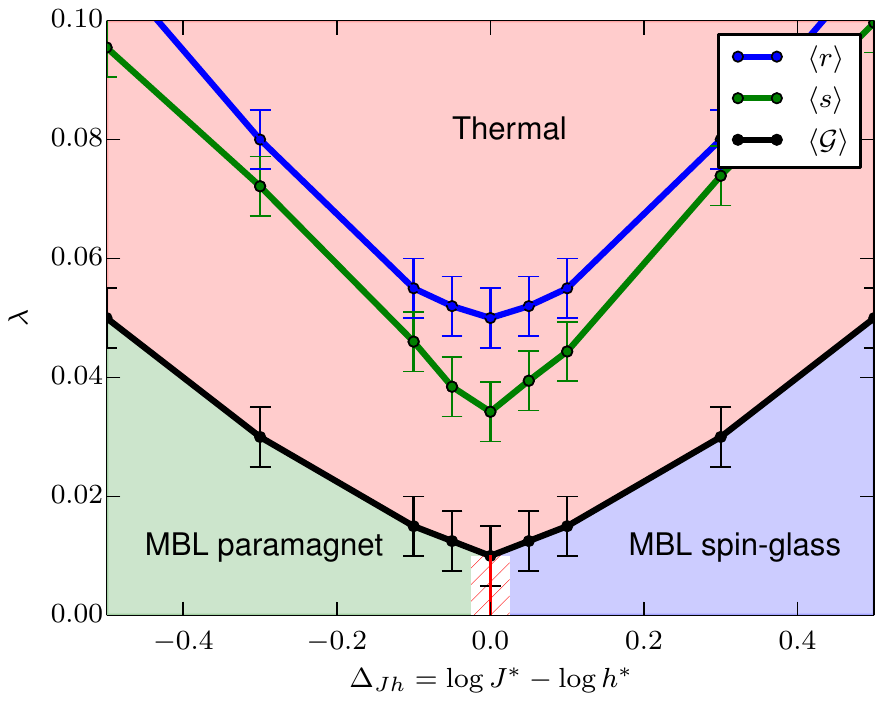}
\caption{Finite size phase boundaries of the disordered self-dual Ising model~\eqref{hamil} in the vicinity of the CP, numerically obtained using three diagnostics discussed in Sec.~\ref{sec:modeldiag}.
If there is a direct MBL PM-to-SG transition for a non-zero interaction strength $\lambda > 0$, it will be at the self-dual $J^*=h^*$ line (red). However, the most sensitive diagnostic, $\langle\mathcal{G}\rangle$, already shows the onset of thermalization at a small value of the dimensional interaction strength, $\lambda_c\sim 1\%$ at the self-dual $J^*=h^*$ line. These phase boundaries are estimated using exact diagonalization on systems of size $L \leq 14$. We also find that the boundaries strongly drift towards towards smaller $\lambda$ as $L$ is increased,  suggesting an instability of the CP for any non-zero $\lambda$ in the infinite size limit. The white shaded box represents this possibility \emph{i.e.} that the thermal phase extends all the way to $\lambda_c=0$, which is suggested by our data and analysis but is inaccessible to finite-size numerics. }
\label{fig:phasediagram}
\end{figure}

This paper is organized as follows. 
In Sec.~\ref{sec:modeldiag}, we study a self-dual TFIM using various diagnostics accessible to many-body exact diagonalization (ED). We find that even for the small system sizes accessible to ED, the CP thermalizes at a tiny interaction strength,  $\lambda_c \sim 1/100$th the size of non-interacting couplings (see phase diagram in Fig.~\ref{fig:phasediagram}), and the size of the critical interaction strength drifts down with increasing system size. 
The small value of $\lambda_c$, together with the incipient thermalization already noticeable at small sizes, strongly suggests a perturbative instability of the CP. We study this further using a novel measure introduced in Sec.~\ref{sec:susceptibility},  the \emph{susceptibility of entanglement} $\chi_S$, which serves as a sensitive probe of the effect of interactions on the entanglement properties of many-body eigenstates. We derive a perturbative expansion for $\chi_S$ in terms of the non-interacting eigenstates  (Sec.~\ref{sec:perturbative}), and use this to elucidate the low-order processes that could destabilize localization at the CP (Sec.~\ref{sec:typicalresonant}). We expect $\chi_S$ to be of independent interest in examining the perturbative effect of interactions in different contexts.
\section{Model and Diagnostics}
\label{sec:modeldiag}
We study a disordered statistically self-dual TFIM in a one dimensional spin-1/2 system of length $L$ with open boundary conditions:
\begin{eqnarray}
    &H = \sum_{i}{J_i \sigma^x_i\sigma^x_{i+1}} + \sum_i{h_i \sigma^z_i} \nn \\
    &+ \lambda \sum_i{(\hm \sigma^z_i \sigma^z_{i+1} + \Jm \sigma^x_i \sigma^x_{i+2})} \equiv H_0 + \lambda V.
\label{hamil}
\end{eqnarray}
The couplings $\{J_i\}  \in [0, \Jm]$ and fields $\{h_i\}\in [0, \hm]$ are drawn independently and randomly from uniform distributions. When $\lambda =0$, the model maps on to Anderson localized non-interacting Majorana fermions via a Jordan Wigner transformation, see App.~\ref{app:JWtransformation} for a review. We appropriately scale terms in the interaction by the strengths of the non-interacting couplings $\Jm$ and $\hm$, which allows us to study a \emph{dimensionless} parameter $\lambda$ that sets the interaction strength while preserving statistical self-duality. As $J^*$ and $h^*$ are tuned to sweep across the phase diagram, we pick $\text{max}[J^*, h^*]=1$, which simply sets an overall scale in the Hamiltonian~\eqref{hamil} because of the manner in which the interaction strength is scaled.  We note that localization is considered truly unstable only if thermalization  happens for \emph{any} strength or configuration of the disordered couplings, and we indeed observe the same qualitative behavior for the analysis presented in the subsequent sections even for ``stronger" power-law disorder.\footnote{However, power-law disorder distributions with large exponents are plagued with strong finite size effects since they have a long tail to very small values; this leads to large separations of scale between local couplings, effectively ``cutting" the chain into smaller pieces in finite-size numerics. Nevertheless, while more challenging to demonstrate numerically, we expect the qualitative arguments and resonant processes identified in this work to also hold for power law disorder distributions that have been considered in various works~\cite{sbrg, vasseurph}}.

The Hamiltonian in Eq.~\eqref{hamil} has $Z_2$ Ising symmetry $P=\prod_i \sigma_i^z$, $[H,P]=0$. 
Deep in the SG phase, $\Jm \gg \hm$, \emph{all} many-body (MB) eigenstates look like pairs of Ising symmetric superposition ``cat" states with $P=\pm 1$ and spins (approximately) aligned randomly along the $x$ axis: $|n_\pm\rangle \approx \frac{1}{\sqrt{2}}\left(|\leftarrow \rightarrow \rightarrow \cdots \leftarrow \rangle \pm |\rightarrow \leftarrow \leftarrow \cdots \rightarrow \rangle\right)$. Domain wall excitations between oppositely oriented spins are localized, and the system shows long-range spin glass order: connected correlation functions of $\sigma^x$ evaluated in $|n_\pm\rangle$ are non-zero for arbitrarily distant correlators, but with random sign~\cite{huse2013localization, kjall2014many}. By contrast, without disorder, eigenstates at any finite temperature have a non-zero density of delocalized domain walls which destroy LRO in one dimension. 
In the opposite limit in the PM, $\hm \gg \Jm$, the eigenstates resemble random product states in the $z$ basis. Far enough from the CP, the localized non-interacting SG and PM phases are stable to weak interactions, thermalizing only at a strong enough interaction strengths $\lambda = \lambda_c > 0$.

Without interactions, the SG-to-PM transition occurs at $\Delta_{Jh} \equiv \log\Jm - \log\hm=0$ due to the Kramers-Wannier duality which maps the PM and SG phases to each other. This transition takes place across the \emph{entire} MB spectrum, and displays the same infinite randomness scaling at all energy densities. 
It has been argued that interactions are irrelevant to the {\it ground-state} transition (where there are no MB resonances)~\cite{fisher1995critical}; we are instead concerned about the fate of the  CP to the addition of interactions at high energy densities.  We study the phase boundary to thermalization for different $\Delta_{Jh}$: $\lambda_c(\Delta_{Jh})$, focusing on $\Delta_{Jh}=0$. The question is whether $\lambda_c(\Delta_{Jh}=0) = 0$, 
so that infinitesimal interactions thermalize the non-interacting CP in the limit of an infinite system. 
We now turn to a numerical exploration of the phase diagram of this model as a function of $\Delta_{Jh}$ and $\lambda$.

\begin{figure}[t]
\includegraphics[scale = 1]{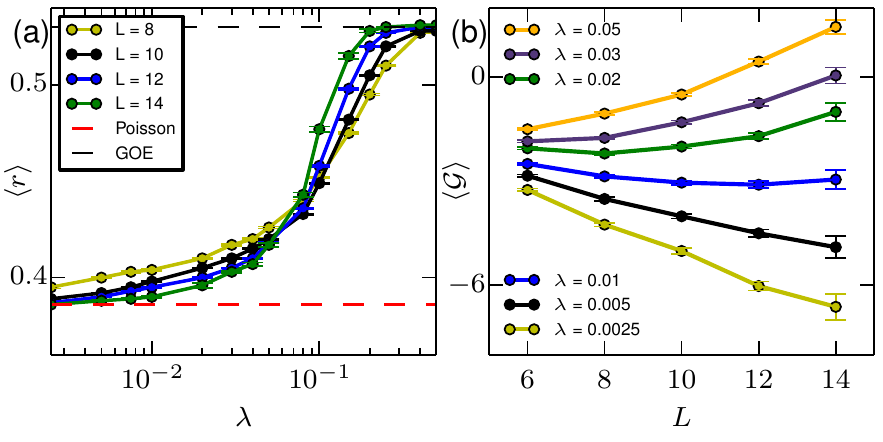}
\caption{Diagnostics of thermalization with increasing interactions for $\Delta_{Jh}=0$. (a) Average $\langle r \rangle$ plotted against $\lambda$. Note the strong drift of the finite-size crossings towards smaller $\lambda$ with increasing system size. (b) $\langle \mathcal{G} \rangle$ as a function of $L$ for different  $\lambda$. Averages are over the middle half of all eigenstates in $3000$ to $300$ independent disorder samples for sizes ranging from $L = 8$ to $L = 14$.}
\label{fig:rG}
\end{figure}

We start by probing the level repulsion in the eigenenergies $\{E_n\}$ within one $Z_2$ symmetry sector $P=1$, via the ratio $ r  = \frac{\min(\delta_n, \delta_{n+1})}{\max(\delta_n, \delta_{n+1})}$ where $\delta_n = E_{n+1} - E_n$~\cite{oganesyan2007localization}. In Fig.~\ref{fig:rG}a, we plot $\langle r \rangle$  vs. $\lambda$ for $\Delta_{Jh}=0$. The average is taken over independent disorder realizations and the middle half of the many-body eigenstates centered around the energy density corresponding to infinite-temperature. The ratio changes from the localized Poisson value, $\langle r \rangle \cong 0.39$, to the thermal value, $\langle r \rangle \cong 0.53$, as $\lambda$ is increased~\cite{atas2013distribution}.
The finite-size crossover to thermalization takes place at the interaction strength $\lambda_c(L_1,L_2)$ where the curves for $L_{1, 2}$ cross.  Note that these crossings are happening at rather small value of the dimensionless interaction, $\lambda_c(12,14) \cong 0.05$ even for the small sizes under study.  
While the sizes are too small to estimate the asymptotic critical interaction strength $\lambda_c$ at large $L$, the locations of the finite-size crossings strongly drift towards smaller $\lambda_c$ with increasing $L$, consistent with the possibility that infinitesimal interactions destabilize the CP.  The finite-size $\lambda_c(L_1,L_2)$ from the crossings between the two largest system sizes for different $\Delta_{Jh}$ are shown in Fig.~\ref{fig:phasediagram} (blue).

\begin{figure}[t]
\includegraphics[scale = 1]{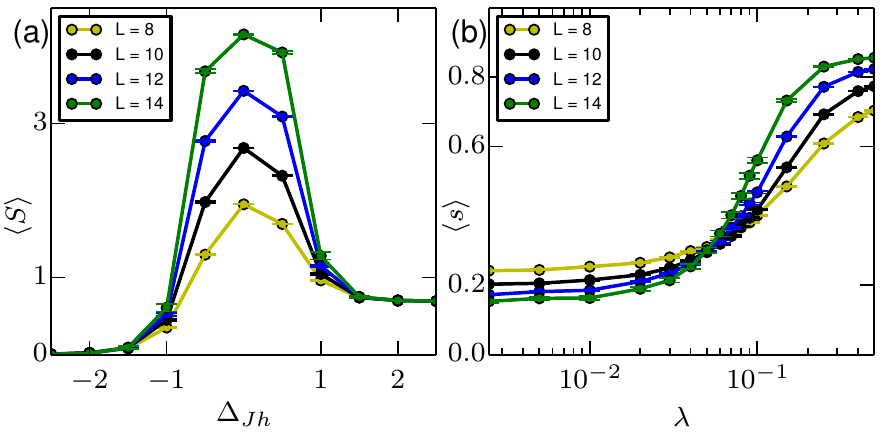}
\caption{Entanglement diagnostics (a) Mean value of the entanglement entropy across a horizontal cut on the phase diagram of Fig.~\ref{fig:phasediagram} for $\lambda = 0.5$. The entanglement entropy clearly changes from $0$ in the PM phase to a volume-law in the thermal phase to $\log 2$ in the spin-glass phase. 
(b) Mean value of the entanglement entropy density $s = S/S_{max}$ across a vertical cut on the phase diagram of Fig.~\ref{fig:phasediagram} for $\Delta_{Jh} = 0$. The entropy density decreases to zero in the critical phase and increases towards 1 in the thermal phase. Averages are over the middle half of all eigenstates in $3000$ to $300$ independent disorder samples for sizes ranging from $L = 8$ to $L = 14$.}
\label{fig:diag}
\end{figure}

A separate diagnostic for the MBL transition is $\mathcal{G} = \log\left(\frac{| \bra{n+1}\widehat{O}\ket{n}| }{\left(E_{n+1} - E_n\right)}\right)$, which probes the ratio of matrix elements to energy gaps for a local operator $\widehat{O}$ in nearby energy eigenstates~\cite{serbyn2015criterion}. The MB energy spacings are exponentially small in $L$; $\widehat{O}$ strongly mixes nearby eigenstates in the thermal phase resulting in $\langle \mathcal{G} \rangle \sim L$, while the mixing is exponentially suppressed in the localized phase resulting in $\langle \mathcal{G} \rangle \sim -L$.
Fig.~\ref{fig:rG}b plots $\langle \mathcal{G} \rangle$ averaged over the middle half of the spectrum for various $\lambda$ and $\Delta_{Jh}=0$ with $\widehat{O} = \sigma^z_{L/2}$. $\lambda_c(L)$ is diagnosed by the change in the sign of the slope of $\langle \mathcal{G}\rangle(L)$. Repeating this for different $\Delta_{Jh}$ gives the finite-size $\lambda_c$ estimates shown in Fig.~\ref{fig:phasediagram} (black).  Note that this diagnostic is more sensitive to thermalization than $\langle r \rangle$ and, among the considered diagnostics, gives the smallest $\lambda_c(\Delta_{Jh})$ finite-size estimates, $\lambda_c(L=14)\simeq 0.01$ at $\Delta_{Jh}=0$. 
We emphasize that $\lambda$ is dimensionless, so it is quite striking that interactions only $1/100$th the value of the couplings is enough to show indications of thermalization even at these small sizes. This small value of $\lambda_c(L)$ and the strong finite-size drifts towards smaller $\lambda_c(L)$ as $L$ is increased, combined with the typical tendency of thermalization to dominate with increasing system size~\cite{devakul2015early, khemani2017critical, doggen2018many}, strongly suggest that interactions are a relevant perturbation to the non-interacting Ising CP at finite energy densities.

Finally, the von Neumann entanglement entropy (EE) of eigenstates is another widely-used diagnostic to probe the MBL transition~\cite{pal2010many, luitz2015many, kjall2014many}. The EE displays a ``volume law" scaling in the thermal phase, an area-law scaling in the MBL phase, and a log scaling in the (putative) critical phase~\cite{refael2004entanglement, devakul2017probability}. Deep in the MBL PM and SG phases, since the eigenstates are product and cat states respectively, the eigenstates EEs are approximately $0$ and $\log 2$, respectively.  
The existence of three phases in this system is clearly detected by the average half-chain eigenstate EEs plotted in Fig.~\ref{fig:diag}a along a path where $\Delta_{Jh}$ is tuned from -4 to 4 for a fixed value of (relatively large) interaction strength $\lambda = 0.5$. Next, to probe the critical behavior, we plot in in Fig.~\ref{fig:diag}b the average entropy density, $S/S_{\textrm{max}}$, as a function of $\lambda$ at $\Delta_{Jh}=0$, where $S_{\textrm{max}} = (L/2)\log 2$ is the maximum EE for a subsystem of size $L/2$. This quantity decreases to $0$ with increasing $L$ in the localized/critical phase, while it saturates towards 1 in the thermal phase. The finite size estimates for $\lambda_c$ are again obtained from where these curves cross for the largest pair of $L$'s, and are shown in Fig.~\ref{fig:phasediagram} (green). This measure gives $\lambda_c(12,14) \cong 0.04$, and hence detects thermalization more sensitively than $\langle r \rangle$ but less sensitively than $\langle \mathcal{G}\rangle$.

\section{Susceptibility of Entanglement}\label{sec:susceptibility}

The incipient thermalization at small $\lambda$ motivates us to introduce a new diagnostic, the susceptibility of entanglement $\chi_S (E, L)$, which perturbatively probes the effect of adding weak interactions on the entanglement of eigenstates at energy density $E/L$ --- specifically, the perturbative relevance of adding interactions may manifest as a divergence in $\chi_S (E,L)$.  

To define $\chi_S(E,L)$, we expand the half-chain EE, $S^{(n)}(\lambda, L)$,  of the $n$-th many-body eigenstate $\ket{n}$ at energy density $E/L$ near interaction strength $\lambda=0$:
\begin{eqnarray}
    &S^{(n)}(\lambda, L) = S_0^{(n)}(L) + \lambda S_1^{(n)}(L) + \frac{\lambda^2}{2} S_2^{(n)}(L) + \cdots,\nn \\
    &\chi_S(E, L) \equiv \langle S_2^{(n)}(L) \rangle_{\textrm{typ}(E)}, 
\label{entropyexpand}
\end{eqnarray}
where $\langle \cdot \rangle_{\textrm{typ}(E)}$ denotes the typical (median) value over samples and eigenstates within small energy density windows of $\delta=0.05$ centered around $E/L$. 
While  $S^{(n)}(\lambda, L)$ can be chosen to be any measure of entanglement of the eigenstate $\ket{n}$, in this work we choose $S^{(n)}$ to be the second Renyi entropy of $\ket{n}$, i.e. $S^{(n)} \equiv  -\log(\mbox{Tr} [\left(\mbox{Tr}_B |n\rangle \langle n|\right)^2)])$,\footnote{Note that $S^{(n)}_j$ in the expansion of Eq.~(\ref{entropyexpand}) should not be confused with the $j$th or $n$th Renyi entropy.} where $B$ is taken to be half the system, and we observe the same qualitative behavior for the von Neumann entropy.
This choice of entropy enables a simple analytical expression of $S^{(n)}_2$ in terms of SP eigenstates (App.~\ref{app:renyiexpand}), allowing us to better elucidate the different contributions to $\chi_S(E)$. It is known that $S^{(n)}_0(L)\sim \log(L)$ for the marginally localized non-interacting CP at $\Delta_{Jh}=0$~\cite{refael2004entanglement, devakul2017probability}.

A key aim of using this susceptibility as a diagnostic is to ``subtract out" the non-interacting contribution to the EE, which overwhelms the value of the EE at the CP at small $\lambda$ at small sizes, for example in Fig.~\ref{fig:diag}b. If interactions are perturbatively irrelevant, we expect the higher order corrections to $S^{(n)}_0(L)$ to be small, and they should not grow with $L$ faster than $S^{(n)}_0(L)$.  On the other hand, if interactions are relevant, this may be manifested in a strong growth of $S^{(n)}_2(L)$ with increasing $L$. Importantly, the behavior of $S^{(n)}_2(L)$ detects a susceptibility towards thermalization even when 
the total second order correction --- obtained by multiplying 
$S^{(n)}_2(L)$ by $\lambda^2/2$ --- may still be tiny (less than one bit) at small $\lambda$ and $L$. Hence this measure serves as a much more sensitive test of an incipient instability. 

\begin{figure}[t!]
\includegraphics[scale = 0.9]{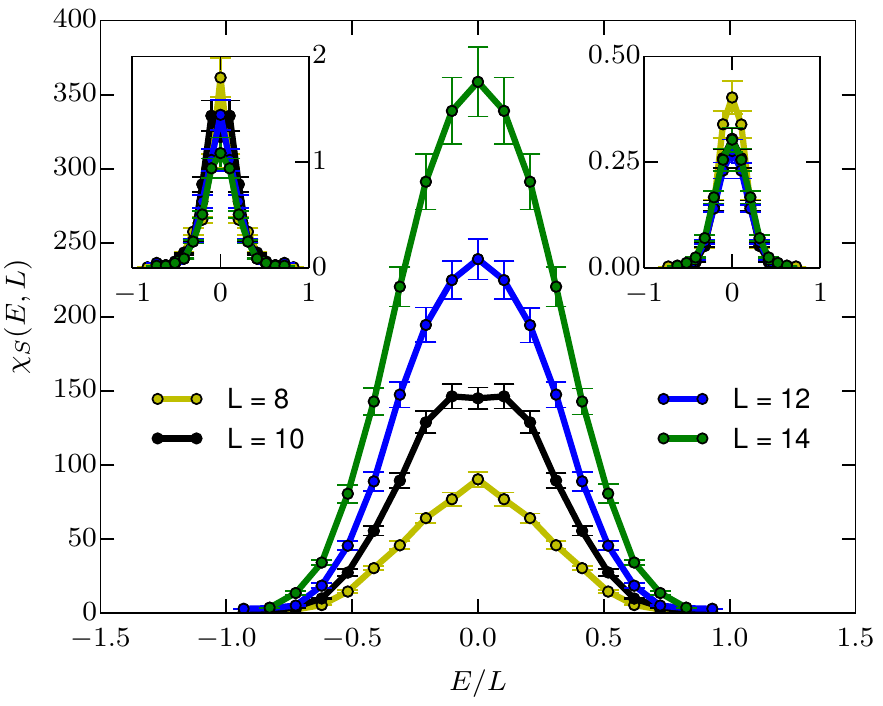}
\caption{$\chi_S(E,L)$, obtained using Eqs.~\eqref{entropyexpand}, \eqref{entropycorrections} as a function of energy density at $\Delta_{Jh} = 0$ (CP, main plot), $\Delta_{Jh} = -1$ (PM, left inset) and $\Delta_{Jh} = +1$ (SG, right inset) with $\epsilon = 10^{-5}$, obtained by taking the median over disorder samples and small energy density windows of $\delta = 0.05$ across the spectrum.
Notice the different scale of the axes of the insets.
The strong increasing trend with $L$ in the middle of the spectrum at the CP points towards a perturbative instability.
Errorbars denote a $2$-percentile interval around the median.}
\label{fig:S2plots}
\end{figure}

To start, we can estimate $S_1^{(n)}(L)$ and $S_2^{(n)}(L)$ numerically using many-body ED and finite-difference derivatives:
\begin{eqnarray}
    &S^{(n)}_1(L; \epsilon) = \frac{S^{(n)}(\epsilon, L) - S^{(n)}(-\epsilon, L)}{2\epsilon} \nn \\
    &S^{(n)}_2(L; \epsilon) = \frac{S^{(n)}(\epsilon, L) + S^{(n)}(-\epsilon, L) - 2 S(0, L)}{\epsilon^2},
\label{entropycorrections}
\end{eqnarray}
where $S^{(n)}(\epsilon, L)$ is the second Renyi entropy of the $n$th MB eigenstate of the Hamiltonian in Eq.~\eqref{hamil} with interaction strength $\lambda=\epsilon$. In doing the finite difference calculations for a given disorder realization, all values of the local fields and Ising couplings are kept the same with only the interaction strength changing.  The expressions above will agree with the ``true" perturbative corrections in the RHS of Eq.~\eqref{entropyexpand} in the limit $\epsilon\rightarrow0$. Instead, when the interaction strength $\epsilon$ is larger than the typical many-body level spacing, we expect (avoided) level crossings in the many-body spectrum of the finite system. In this regime, the interacting MB eigenstate $\ket{n (\lambda)}$ can no longer be interpreted as being perturbatively related to the non-interacting MB eigenstate $\ket{n (\lambda=0)}$. Hence, to probe the regime where non-degenerate perturbation theory is valid, we use $\epsilon \leq 10^{-5}$, which we numerically find is smaller than the typical many-body level spacing for all the system sizes $L \leq 14$.

Strikingly, even for this tiny value of $\epsilon=10^{-5}$ which is three orders of magnitude smaller than the smallest estimate of $\lambda_c\approx 0.01$ at these sizes, we find that $\chi_S(E, L)$ shows a strong increasing trend with $L$ for finite energy densities at the CP ($\Delta_{Jh}=0$), strongly indicative of a perturbative instability of localization (Fig.~\ref{fig:S2plots}).
In contrast, far from the CP in the SG/PM phases, the typical value of $\chi_S(E, L)$ is two orders of magnitude smaller and there is no strong trend with $L$ (shown in the insets of Fig.~\ref{fig:S2plots}), consistent with the stability of localization in these regimes. Interestingly, we continue to see this strong growth of $\chi_S(E, L)$ at the CP even for much larger values of $\epsilon$ that are in the non-perturbative regime (but still well below the finite-size $\lambda_c$ estimates). 

Finally, we note that we have defined $\chi_S(E, L)$ in terms of the second order correction to the entropy rather than the first. This is because we find that $S^{(n)}_1(L)$ appears with a random fluctuating sign, and its typical value is thus small for large system sizes, while $S^{(n)}_2(L)$ typically gives a strong positive correction. Even upon taking an absolute value, the typical value of $|S_1^{(n)}(L)|$ is much smaller than $S^{(n)}_2(L)$ with a much weaker growth with system size.
This is also because, $S^{(n)}_1(L)$ admits an expansion as a sum of contributions with fluctuating signs that give a suppressed correction, as we show in App.~\ref{app:renyiexpand}. 
\section{Perturbative expression of \texorpdfstring{$\boldsymbol{\chi_S(E, L)}$}{}}\label{sec:perturbative}
Given that a strong growth of $\chi_S(E, L)$ with system size is already visible in the regime of non-degenerate perturbation theory, we now try to understand its behavior using a perturbative expansion of the second R\'{e}nyi entropy $S^{(n)}(\lambda, L)$ derived using second-order perturbation theory on the non-interacting model $H_0$ (App.~\ref{app:renyiexpand}). The many-body eigenstates and eigenvalues of $H_0$ (obtained by `filling' SP orbitals in the fermionic language) are denoted $|\psi^0_n\rangle$ and $E_n^0$ respectively. The expression for $S^{(n)}_2$ obtained by perturbing $|\psi^0_n\rangle$ to second order is of the form (see Eq.~(\ref{eq:S2expand}))
\begin{equation}
S^{(n)}_2 = \sumal{k \neq n}{}{c_{kn} g_{kn}^2} + \sumal{k \neq l \neq n}{}{d_{k l n} g_{kn} g_{ln}} + \sumal{k \neq n}{}{e_{kn} \alpha_{kn}},
\label{eq:S2form}
\end{equation}
where the functional dependence of the quantities on $L$ is implicit.
The sums in Eq.~(\ref{eq:S2form}) run over all many-body eigenstates $n$ and $k$ of $H_0$, $c_{kn}$, $d_{kln}$, and $e_{kn}$ are $\mathcal{O}(1)$ numbers that are related to the properties of the many-body wavefunctions (see Eqs.~\eqref{eq:ckndefn}-(\ref{eq:ekndefn}) for their definitions), and $g_{kn}$ and $\alpha_{kn}$ involve ratios of matrix elements of the interaction and energy denominators that can lead to large values of $S^{(n)}_2$ (see Eq.~(\ref{eq:ratios}) for their definitions). Since the full expressions for $S_{1/2}^{(n)}$ are quite complicated, we are not aware of an efficient algorithm to compute all terms in the perturbative correction and \emph{exact} numerics are still limited to small system sizes accessible to many-body ED. Hence, in the rest of this work, we use numerical observations to perform various approximations that allow us to speculate on the behavior of $\chi_S(E, L)$ at larger system sizes than those accessible to exact numerics.

First, note that the $g_{kn}$'s and $\alpha_{kn}$'s in Eq.~\eqref{eq:S2form} typically appear with fluctuating signs. If one neglects correlations between different terms, we expect the second and third sums to not result in large contributions to $S^{(n)}_2$.
Indeed, we numerically observe that the typical value of $S_2^{(n)}$ receives its dominant contribution from the first sum (involving $g_{kn}^2$) when $n$ is a highly excited many-body eigenstate, thereby allowing us to approximate
\begin{equation}
     \chi_S(E, L) = \langle S^{(n)}_2(L) \rangle_{\textrm{typ(E)}} \sim \langle \sum_{k\neq n} c_{kn} g_{kn}^2\rangle_{\textrm{typ(E)}}.
     \label{eq:S2defn}
\end{equation}
This approximation is \emph{not} valid when restricted to the ground state, for which we find that the different terms in the expansion conspire and cancel to give a small $S^{(n)}_2$ without a strong system size dependence, consistent with arguments for the perturbative stability of the ground state at and away from the CP~\cite{fisher1995critical}. 

Next, note that $g_{kn}$ is defined as (see Eq.~(\ref{eq:ratios}))
\begin{align}
g_{kn} \equiv \frac{V_{kn}}{E_{kn}} \equiv \frac{\bra{\psi^0_k} V \ket{\psi^0_n}}{E^0_n - E^0_k},
\label{eq:cg1}
\end{align}
where $V_{kn}$ is the matrix element of $V$ between the MB eigenstates of $H_0$, $|\psi^0_{n/k}\rangle$, with MB energies $E^0_{n/k}$. Analogous to the diagnostic $\langle\mathcal{G}\rangle$ studied earlier, $g_{kn}$ probes the ability of $V$ to generate resonances between the eigenstates of $H_0$, and can systematically grow or shrink with $L$. When written in terms of fermions, $V$ is quartic and fermion parity preserving (see Eqs.~(\ref{eq:interactionquartic}) and (\ref{eq:interactionaway})).
Hence $V_{kn}$ vanishes unless $\ket{\psi^0_k}$ and $\ket{\psi^0_n}$ differ in the occupation of two or four single-particle orbitals, and we refer to these as ``two-orbital" and ``four-orbital" processes respectively. This allows for an efficient polynomial in $L$ computation of all $g_{kn}$ for a given $n$ (App.~\ref{sec:matrixelements}). 

\begin{figure*}[t]
\includegraphics[scale = 1]{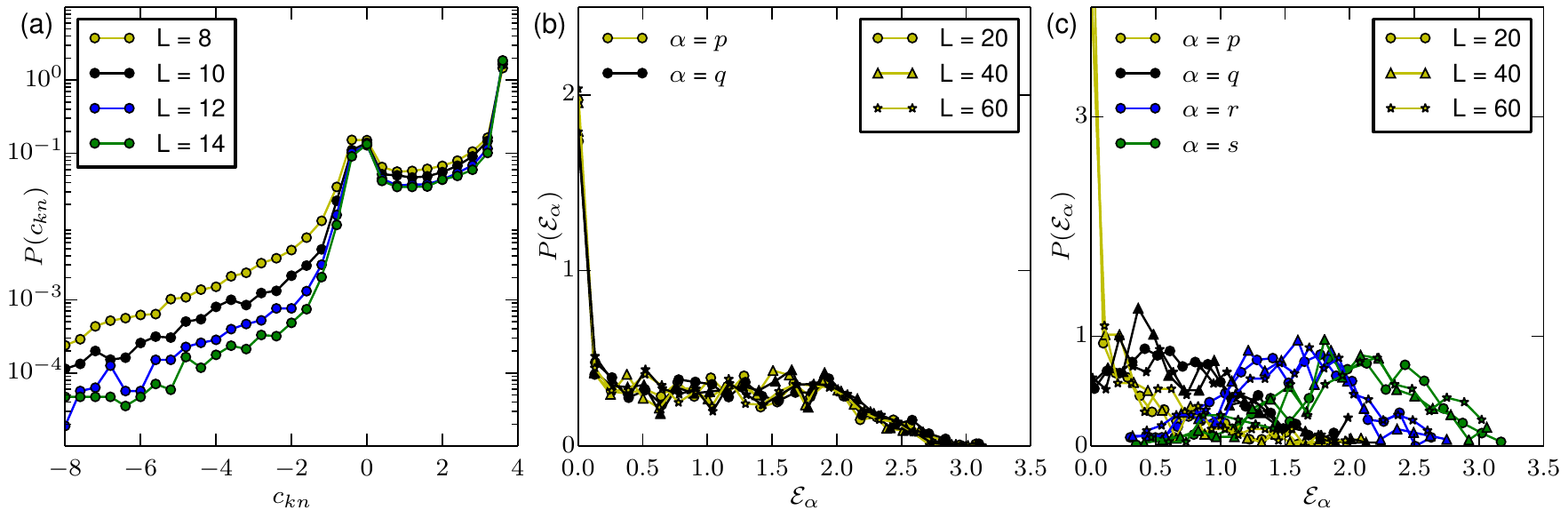}
\caption{(a) Bimodal distribution of $c_{kn}$ for randomly sampled infinite temperature states (b-c) Distribution of SP energies of orbitals involved in strongly resonant ($g_{kn}^2 > 100$) (b) two orbital processes $(\mE_p < \mE_q)$ (c) four orbitals processes $(\mE_p < \mE_q < \mE_r < \mE_s)$. Data is obtained by sampling 100 random processes for each of 30-60 many-body eigenstates for 80-250 disorder realizations for various system sizes, and choosing the strongly resonant processes ($g_{kn}^2 > 100$).}
\label{fig:42processes}
\end{figure*}

Finally, we observe that $c_{kn}$ is an $\mathcal{O}(1)$ number (defined in Eq.~\eqref{eq:ckndefn}) that is bounded as $-8 \leq c_{kn} \leq 4$.
Unfortunately, we have not been able to obtain a similarly efficient numerical method for computing $c_{kn}$. However, note that: (i) $c_{kn}$ is an $\mathcal{O}(1)$ number, so it cannot be the source of any divergence in $\chi_S(E, L)$ with system size and (ii) we have found that the primary role of $c_{kn}$ is to enforce the condition that the entanglement only receives a substantial contribution from processes that toggle the occupation of SP orbitals with weight on opposite or both sides of the entanglement cut. The latter statement has been numerically verified, and we observe that $c_{kn}$ is strongly bimodal, peaked at 0 and 4 (Fig.~\ref{fig:42processes}a), and these peaks respectively correspond to whether or not the orbitals involved straddle the entanglement cut.
This is consistent with the fact that resonances between orbitals localized on the same side of the entanglement cut cannot contribute to any extra entanglement.
As a result, despite the fact that $c_{kn}$ can be negative, the sum in the first term of Eq.~\eqref{eq:S2form} is typically strongly positive. In the next section, we examine the behavior of $g_{kn}$ to identify the processes that could lead to a growth of $\chi_S(E, L)$. 
\section{Resonant Processes in the Non-interacting Limit}\label{sec:typicalresonant}
We begin by noting that the distributions of $g^2_{kn}$ for both two-orbital and four-orbital processes are broad on a log-scale, so that the behavior of the typical value of the sum in Eq.~(\ref{eq:S2defn}) is captured by the behavior of the typical value of the \emph{maximum} $c_{kn} g^2_{kn}$: 
\begin{equation}
\chi_S(E, L) \sim \langle \max_{k\neq n}{\left(c_{kn} g^2_{kn}\right)} \rangle_{\textrm{typ}(E)}.
\label{eq:Gdefn}
\end{equation}
for highly excited $n$. We have numerically verified that this approximation captures the dominant contribution to $\chi_S(E, L)$ and tracks its dependence on $L$ for the system sizes accessible to ED for which we can explicitly compute $c_{kn}$. Using this approximation for $\chi_S(E, L)$, we now elucidate the non-interacting processes that could lead to a perturbative instability of the CP for excited states at larger sizes than those accessible to many-body ED. 
Eq.~(\ref{eq:Gdefn}) implies that, for a highly-excited eigenstate $\ket{n}$, the susceptibility of entanglement is dominated by the  \textbf{most resonant} two/four orbital process relative to $|\psi_n^0\rangle$, with orbitals straddling the entanglement cut. 
Perturbative stability is determined by the scaling of $\chi_S(E, L)$ with system size $L$, so we will be concerned with the scaling of the typical most resonant processes that straddle the cut. We find that these diverge with $L$ for highly excited states at the CP and, for the sizes we have been able to study, we find that the four-orbital processes are dominant over the two-orbital ones and capture almost the entire growth of $\chi_S(E, L)$. Note that this sensitivity to the typical most resonant processes is qualitatively distinct from measures such as $\langle \mathcal{G}\rangle$, which are instead analogous to the typical $g_{nk}$ which is exponentially decaying with $L$ at and away from the CP.

We now identify possible sources of the instability of the non-interacting Ising CP at $\Delta_{Jh}=0$.
As reviewed in App.~\ref{app:nonint}, the SP eigenstates at the CP have a diverging localization length as the SP energy $\mE \rightarrow 0$; these orbitals are instead \emph{stretched} exponentially localized: $|\psi_\alpha(x)| \sim e^{-\sqrt{|(x-x_\alpha)|/\xi}}$, where $\psi_\alpha(x)$ denotes the $\alpha$-th SP wavefunction. \footnote{More precisely, the orbitals at SP energy $\mE$ are stretched exponentially localized on a length scale $\zeta(\mE)$, with a crossover to exponential tails on longer length scales, and $\zeta(\mE) \rightarrow L$ as $\mE \rightarrow0$.} Likewise, the density of states diverges near $\mE \rightarrow 0$, so that the typical energy spacing between these stretched exponentially localized orbitals \emph{also} scales as a stretched exponential: $\delta \mE \sim  e^{-\sqrt{sL}}$. By contrast, away from $\mE=0$ at the CP (or away from the CP, $\Delta_{Jh}\neq0$, at any $\mE$), the SP orbitals are exponentially localized, with the energy spacings only fall off as laws in $L$. The presence of extended orbitals near zero energy can mediate two and four orbital resonances, as we explain next. 

We consider first the two-orbital processes which allow for a more transparent explanation, even though the dominant contribution comes from four-orbital terms. The relevant two-orbital processes are those in which two stretched exponentially localized states, $p$ and $q$, with $\mE_{p,q}\simeq 0$ and $\Delta_{pq}=|\mE_p \pm \mE_q| \sim  e^{-\sqrt{R_{pq}/\Delta}}$, have their occupations toggled between $|\psi_n^0\rangle$ and $|\psi_k^0\rangle$. Since $V$ is a sum of local operators, the matrix element $V_{kn} \sim e^{-\sqrt{R_{pq}/\Xi}}$ is itself stretched exponentially decaying. The localization centers between these states are typically separated by $R_{pq}\sim L$. Thus, $g_{kn}= V_{kn}/\Delta_{pq}$ can be divergent depending on the relative sizes of $\Delta$ and $\Xi$. As we discuss in App.~\ref{app:nonint}, the distributions of the stretched-exponential forms of the end-end correlations and energy gaps for the ground state at the CP were derived in Ref.~\cite{fisher1998distributions}, and the scales of those were shown to be comparable.
We similarly expect a comparable scaling for $\Xi$ and $\Delta$ for the matrix elements of local operators and gaps involving extended SP orbitals near $\mE=0$, so that there is a finite probability for a resonance with a divergent $g_{kn}$. Fig~\ref{fig:42processes}b samples the energy distributions of the SP orbitals involved in strongly resonant processes ($g_{kn}^2 > 100$) for randomly chosen excited states across random samples, confirming that resonances are due to two extended orbitals close to $\mE = 0$. By contrast, in typical processes involving exponentially localized orbitals, the matrix element $V_{kn}$ typically decays exponentially with $R_{pq}$, while the energy difference only decays as a power law so that there is typically no resonance.

Note that we are able to go to much larger sizes of $L\sim 60$ because of the polynomial in $L$ scaling of the time for computing $g_{kn}$. However, the random sampling of resonant states is not the same as sampling $\langle\text{max}_{k\neq n} g_{kn}^2\rangle_{\text{typ}(E)}$, which is more computationally intensive. However, the random sampling still qualitatively illustrates the types of processes that can lead to resonances at the CP. Note also that the two body processes considered here are first-order in the interaction strength, and they probe a qualitatively distinct process from the second-order processes analysed in Ref.~[\onlinecite{nandkishore2014marginal}], which also studied the stability of marginal Anderson localized systems. Indeed, the Ising CP is stable to the processes considered in Ref.~\cite{nandkishore2014marginal} (App.~\ref{app:NP}).

Next, we consider four-orbital processes involving orbitals $p, q, r, s$ with energies $\mE_{p, q, r,s}$. The energy denominator thus reads $\Delta_{pqrs}=|\mE_p \pm \mE_q \pm \mE_r \pm \mE_s|$, where the signs depend on the occupations of the four orbitals in $\ket{\psi^{0}_n}$, as we explain in App.~\ref{sec:matrixelements}.
As shown in Fig~\ref{fig:42processes}c, typical resonant four-orbital processes at these sizes involve 1-2 extended orbitals near SP energy $\mE = 0$, and 2-3 exponentially localized orbitals with energies away from zero.
The wavefunctions of the localized orbitals overlap with the stretched exponential wavefunctions in order to obtain a substantial matrix element mediated by the delocalized orbital(s).
As the system size is increased and the SP spectrum ``fills in", there are more delocalized orbitals near $\mE \approx 0$, and the resonant processes may involve more extended orbitals. At the sizes amenable to our analysis, we find that four-orbital processes are  dominant over two-orbital ones at the CP. Away from the CP, all SP orbitals are exponentially localized and four-orbital resonances are highly suppresed. 

As a cautionary remark, we note that while we have focused on identifying processes at the CP that can give rise to resonant $g_{kn}$'s, this alone is not enough to argue for a diverging $\chi_S(E, L)$, and the $c_{kn}$'s in Eq.~\eqref{eq:Gdefn} do play a significant role.
It is important that the most resonant processes at the CP typically involve one or more extended orbitals with weight on both sides of the entanglement cut, and hence also give a positive $c_{kn}$ and contribute to $S_2^{(n)}$ via Eq.~\eqref{eq:Gdefn}. In contrast, we find that the typical most resonant processes deep in the PM/SG phases, obtained via flipping a few rare resonant SP orbitals, have a vanishingly small $c_{kn} \approx 0$ and do not contribute substantially to $S_2^{(n)}$. If, instead, we consider the RHS of Eq.~\eqref{eq:Gdefn} which includes $c_{kn}$ then, in the PM/SG phases, the  resonances that contribute typically involve localized orbitals within an $\mathcal{O}(1)$ distance of the entanglement cut and there is no strong system size dependence, consistent with the perturbative stability of the PM/SG phases to interactions.
Likewise, even though the presence of extended SP orbitals at the CP may, in principle, also destabilize the ground state of the CP, we do not see any signs of this in practice because (i) the energy denominators in $g_{kn}$ when $n$ is the ground state always involve the \emph{sum} of SP eigenenergies (as opposed to sums and differences in excited states) and are less likely to be resonant and    
(ii) when $n$ is the ground state, the approximation in Eq.~\eqref{eq:S2defn} is not a good one and one needs to consider the full expression Eq.~\eqref{eq:S2form} on account of various correlations and cancellations between these terms.

In summary, it required a conspiracy of many factors to conclude that the typical \emph{most resonant} two/four body process relative to $|\psi_n^0\rangle$ --- which involves changing the occupation of one or more extended SP orbitals at the CP --- is able to dominantly capture the behavior of $S^{(n)}_2$ for excited states at the CP.
We also note that while Eq.~\eqref{eq:S2defn} only considers a single MB state $k$ relative to $n$, we are not suggesting that a single pair of resonant MB states is sufficient to thermalize the system at large sizes. Rather, the resonance between the many-body eigenstates $(n,k)$ and the corresponding strong growth of $S_2^{(n)}(L)$ with $L$ is only capturing the \emph{perturbative beginnings of an incipient instability} towards thermalization. We remind the reader that, at the sizes amenable to ED, the product $\lambda^2 S_2^{(n)}(L)/2$ is still much less than one bit of entanglement, and hence the second order correction to $S_0^{(n)}(L)$ is still very small in absolute terms. However, the strong trend of growth of $S_2^{(n)}(L)$ with $L$ strongly suggests that the nascent signature in $\chi_S(E, L)$ may lead to a thermalizing cascade across the entire MB energy spectrum as higher order processes and larger sizes are considered.

\section{Concluding Remarks}\label{sec:conclusions}
We have presented a study of the stability of the marginally Anderson localized spin-glass to paramagnet critical point in a disordered transverse-field Ising model. Within many-body ED, we obtain a finite-size estimate for the critical (dimensionless) interaction strength, $\lambda_c(L)\sim 1\%$. This already tiny value of $\lambda_c(L)$ even for modest system sizes $L \leq 14$, coupled with a drift towards smaller $\lambda_c$ with increasing $L$, point to a perturbative instability of the CP to the addition of interactions \emph{i.e.} $\lambda_c=0$ in the asymptotic infinite size limit. 

We introduced a new measure, the ``susceptibility of entanglement" $\chi_S(E, L)$, which perturbatively probes the effect of adding interactions on the entanglement of non-interacting eigenstates. This serves as a much more sensitive probe of an incipient instability, and shows a strong divergence with $L$ at the CP even when estimated to leading order in an arbitrarily weak interaction. Using a perturbative expansion for $\chi_S(E, L)$, we related the susceptibility to the ratio of matrix elements and energy differences in the non-interacting problem, and identified that resonances mediated by extended single-particle states at the CP are the leading order processes contributing to the growth of $\chi_S(E, L)$ with system size $L$. At these sizes and interaction strengths, the absolute (in magnitude) correction to the non-interacting entanglement is still very small; but the strong divergence of $\chi_S(L)$ with $L$ points to the perturbative beginnings of an incipient instability that could lead to full thermalization across the MB energy spectrum as higher-order processes and larger sizes are considered. 

An important point is that a divergence in $\chi_S(E, L)$ may be caused even if the addition of interactions leads to a discontinuous change in the critical eigenstate entanglement entropy, for example from $S \sim c_1 \log L$ in the non-interacting limit to $S \sim c_2 \log L$ in the interacting problem. While such a change does not correspond to thermalization, it still points to a relevance of interactions at the non-interacting infinite-randomness critical point. This (weaker) effect seems unlikely for highly-excited eigenstates at the CP, which already show signatures of thermalization for small $\lambda$ in finite size ED studies. However, it would be interesting to examine whether interactions might be relevant in this weaker sense for the ground state phase transition. Indeed, revisiting various strong randomness RG treatments of the Ising transition~\cite{pekker2014hilbert, vosk2014dynamical} - which explicitly ignore the possibility that the interactions are relevant - with these considerations in mind is an important direction for future work.

While this paper has focused on a perturbative instability of the CP to the addition of interactions, we note that the system may also be subject to non-perturbative instabilities on account of the diverging SP localization length near $\mE=0$~\cite{de2017stability}. These effects, if present, would be subdominant to the perturbative processes, and would only be visible at asymptotically larger sizes than those considered here. However, these may play a dominant role once the system is perturbed slightly away from criticality, in which case the non-interacting localization length $\xi(\mE)$ remains finite but may get very large as $\mE \rightarrow 0$. Due to the energy dependence of the localization length, this case is not directly covered by the arguments in Ref.~\cite{de2017stability} which assume a uniform $\xi$ across the SP spectrum and predict an instability once $\xi$ exceeds an $\mathcal{O}(1)$ threshold. Instead, this requires a more nuanced analysis along the lines of the recent study in Ref.~\cite{crowley_unstable}, which allowed for a distribution in $\xi(\mE)$.   

We reiterate that an \emph{instability} of the marginally Anderson localized CP corresponds to (asymptotic) thermalization in the presence of interactions, independent of the details or strength of the disorder configurations. However, stronger disorder, for instance generated by power law distributions, is much harder to analyze numerically because of strong finite-size effects. The power-law disorder generates tails to very small couplings, which effectively decouples the system into smaller pieces and can make the system look more localized than it is. We have repeated our analysis for power-law disorder and broadly found qualitatively similar behavior. However, even for (not too strongly disordered) power-law distributions with exponent $2$, we find that the many-body energy spacings become small enough that a perturbative estimate of $S_2(L;\epsilon)$ for small $\epsilon$ runs into machine precision issues. Hence prior numerical studies of MBL-to-MBL phase transitions that use power-law distributions with large exponents~\cite{sbrg, vasseurph} should be interpreted with caution.  

Finally, while we have focused on the PM to SG phase transition in a disordered TFIM, our considerations are expected to apply more broadly to various putative MBL-to-MBL phase transitions with a marginally Anderson localized non-interacting counterpart. Some of these transitions, for example those between the ``$\pi$ spin-glass" (or discrete time crystal) phase and various paramagnetic phases in a driven Ising model~\cite{khemani2016phase} are in the infinite randomness universality class as the disordered TFIM~\cite{Berdanier_pnas} and will be subject to a similar instability once interactions are added. Separately, Ref.~\cite{vasseurph} considered a ``particle-hole" symmetric disordered XXZ chain; they found that the addition of interactions to the non-interacting (critical) disordered XX chain could produce a localized spin glass phase with spontaneously broken particle-hole symmetry. In the fermion language, the non-interacting limit of their model corresponds to two decoupled and critical Majorana chains with an additional reflection symmetry, which enabled the possibility of localization via symmetry breaking on adding interactions --- a possibility that is absent in the critical random TFIM which corresponds to a single critical Majorana chain. The disordered XXZ model represents a particle-hole symmetric slice through a broader class of disordered XYZ spin chains with localized spin-glass phases pointing along the $x/y/z$ directions~\cite{sbrg}. If we consider the phase diagram in this broader parameter space, our considerations are expected to apply to transitions between glassy phases ordered in different directions. This model was studied via  ``spectrum-bifurcation" RG in Ref.~\cite{sbrg}, and this RG scheme again did not include the possibility of instabilities at the critical lines.
Ref.~\cite{sbrg} also presented an ED analysis, and found that they needed power-law disorder with a large exponent (greater than 4) to prevent thermalization at the critical lines --- but this regime is not trustworthy for the small sizes amenable to their ED analysis, as discussed earlier. 

In all, our work adds to the growing body of work on non-equilibrium quantum criticality, and addresses long-standing open questions about the nature and stability of MBL-to-MBL phase transitions.  We expect $\chi_S(E, L)$ may be of independent interest for studying the effect of interactions in a variety of settings, for example to probe the existence of many-body mobility edges, another major outstanding question in the literature. While we had to rely on various approximations to study $\chi_S(E, L)$, it would also be interesting to see if more exact methods could be developed to study this quantity at larger sizes, given its perturbative nature. Separately, 
different techniques such as matrix product state based methods \cite{khemani2016obtaining, pollmann2016efficient, yu2017finding, devakul2017obtaining}, or numerical linked cluster expansions~\cite{devakul2015early}, or machine learning techniques~\cite{schindler2017probing, doggen2018many, venderley2018machine, van2018bloch} may prove useful for more large scale analyses.

\vspace{3pt}
\noindent \emph{Note Added---} During the completion of this work, we became aware of complementary work on the presence of intervening thermal phases between MBL transitions which will appear in the same arXiv posting~\cite{Sahay}.

\vspace{3pt}
\noindent \emph{Acknowledgements---} We thank Subroto Mukerjee for an initial collaboration, and Trithep Devakul, Chris Laumann,  Siddharth Parameswaran, Shivaji Sondhi and Romain Vasseur for helpful discussions. This work was supported in part by the DARPA DRINQS program.

\appendix
\section{Jordan-Wigner Transformations of Eq.~(\ref{hamil})}\label{app:JWtransformation}
In this appendix, we perform a Jordan-Wigner transformation on the Hamiltonian Eq.~(\ref{hamil}) with open boundary conditions. 
We split the Hamiltonian of Eq.~(\ref{hamil}) as
\begin{equation}
    H = H_0 + \lambda V
\end{equation},
where $H_0$ is the disordered transverse field Ising model 
\begin{equation}
    H_0 = \sumal{j = 1}{L}{h_j\sigma^z_j} + \sumal{j = 1}{L-1}{J_j \sigma^x_j \sigma^x_{j+1}},
\label{eq:disorderedtfim}
\end{equation}
and $V$ is the Kramers-Wannier self-dual interaction that reads
\begin{equation}
    V =  \sumal{j = 1}{L-2}{(\hm\sigma^z_j \sigma^z_{j+1} + \Jm\sigma^x_j \sigma^x_{j+2})}.
\label{eq:interaction}
\end{equation}
$H$ has a $Z_2$ parity symmetry, defined by the operator
\begin{equation}
    P = \prodal{j = 1}{L}{\sigma^z_j},\;\;\; \left[P, H_0\right] = 0, \;\;\left[P, V\right] = 0.
\label{eq:paritysym}
\end{equation}
To perform the Jordan-Wigner transformation to fermions with creation and annhilation operators $c_j$ and $c_j^\dagger$, we apply the transformations:
\begin{eqnarray}
\sigma^+_j &=& (-1)^{\sum_{k<j}{n_k}} c_j^\dagger \nn \\
\sigma^-_j &=& (-1)^{\sum_{k<j}{n_k}} c_j \nn \\
\sigma^z_j &=& 2 c_j^\dagger c_j - 1 = -(-1)^{n_j} 
\label{jw}
\end{eqnarray}
where $n_j = c^\dagger_j c_j$ and $\sigma^\pm_j = \sigma^x_j \pm i \sigma^y_j$, after which $H_0$ maps onto a superconducting Hamiltonian
\begin{equation}
    H_0 = \sumal{j = 1}{L-1}{\left(h_j (2 n_j -1) + J_j (c_j^\dagger c_{j+1} + c_j^\dagger c_{j+1}^\dagger + \textrm{h.c.})\right)},
\end{equation}
and the interaction $V$ maps onto
\begin{eqnarray}
    V &= \sumal{j = 1}{L-1}{( \hm \left(2 n_j - 1\right)\left(2 n_{j+1} - 1\right)} \nn \\
    &+ \Jm (-1)^{n_{j+1}} (c_j^\dagger c_{j+2} + c_j^\dagger c_{j+2}^\dagger + \textrm{h.c.}) ).
\label{eq:interactionquartic}
\end{eqnarray}
The expression for $V$ can be simplified using the property $(-1)^{n_j} = 1 - 2 n_j$.
Furthermore, under the Jordan-Wigner transformation of Eq.~(\ref{jw}), the parity operator $P$ of Eq.~(\ref{eq:paritysym}) maps onto
\begin{equation}
    P = (-1)^L \times (-1)^{\sumal{j = 1}{L}{n_j}}
\end{equation}
Since a superconducting Hamiltonian is better expressed in terms of Majorana fermions, we apply a second set of substitutions
\begin{eqnarray}
c_j^\dagger &=& \frac{1}{2}(\chi_{2j-1}+i \chi_{2j}) \nn  \\
c_j &=& \frac{1}{2}(\chi_{2j - 1} - i \chi_{2j})
\end{eqnarray}
where the $\chi_{2j-1}$'s and $\chi_{2j}$'s are Majorana fermions that obey the commutation relations
\begin{equation}
    \{\chi_a, \chi_b\} = 2\delta_{ab}.
\end{equation}
In terms of the Majorana fermions, the fermion number and parity operators read
\begin{equation}
    n_j =  \frac{1}{2}(1 - i \chi_{2j - 1} \chi_{2j}),\;\; P = (-i)^L \prodal{l = 1}{2L}{\chi_l},
\end{equation}
and $H_0$ and $V$ simplify to
\begin{eqnarray}
    &H_0 = - i\sumal{j = 1}{L-1}{ (h_j \chi_{2j-1}\chi_{2j} + J_j \chi_{2j+1}\chi_{2j})} \label{free} \\
    &V =  -\sumal{j = 1}{L-2}{(\hm \chi_{2j-1}\chi_{2j}\chi_{2j+1}\chi_{2j+2}} \nn \\
    &{+ \Jm\chi_{2j}\chi_{2j+1}\chi_{2j+2}\chi_{2j+3})}\label{eq:interactionaway}
\end{eqnarray}
At the critical point (when $\hm = \Jm \equiv C$), $V$ can be written more elegantly as 
\begin{equation}
    V = C \sumal{j = 1}{2L - 3}{\chi_j\chi_{j+1}\chi_{j+2}\chi_{j+3}},
\label{interaction}
\end{equation}
%
%

%
%
%
\section{Perturbative expansion of the second R\'{e}nyi entropy}\label{app:renyiexpand}
In this appendix, we compute the expression for the susceptibility of entanglement entropy in terms of matrix elements of the non-interacting problem.
For simplicity, we use choose the second R\'{e}nyi entropy to be the entanglement entropy. That is,
\begin{equation}
    S(\lambda, L) \equiv -\log\left({\rm Tr }\ \rho_{\rm red}^2\right)    
\label{Sdefn}
\end{equation}
is the entanglement entropy of a particular eigenstate at an interaction strength $\lambda$ and system size $L$.
We expand $S(\lambda, L)$ in powers of $\lambda$ as shown in Eq.~(\ref{entropyexpand}).
\begin{equation}
    S(\lambda, L) = S_0(L) + \lambda S_1(L) + \frac{\lambda^2}{2} S_2(L) + \dots
\end{equation}
where $S_0(L)$ is the entanglement entropy in the non-interacting limit and $S_1(L)$ and $S_2(L)$ are the first and second derivatives of $S(\lambda, L)$.
To compute $S_1(L)$ and $S_2(L)$ perturbatively, we start with a non-interacting many-body wavefunction $\ket{\psi^0_n}$, and write the perturbed wavefunction $\ket{\psi_n}$ up to $\mo\left(\lambda^2\right)$ as
\begin{eqnarray}
    &\ket{\psi_n} = \left(1 - \frac{\lambda^2}{2} \sumal{k\neq n}{}{|g_{kn}|^2}\right)\ket{\psi^0_n} + \lambda \sumal{k\neq n}{}{g_{kn} \ket{\psi^0_k}} \nn \\
    &+ \lambda^2 \sumal{k \neq n}{}{\alpha_{kn} \ket{\psi^0_k}} + \mathcal{O}\left(\lambda^3\right),
\label{wfperturbed}
\end{eqnarray}
where $\{\ket{\psi^0_k}\}$ is the set of unperturbed non-interacting many-body wavefunctions and 
\begin{eqnarray}
    &g_{kn} \equiv \frac{\bra{\psi^0_k} V \ket{\psi^0_n}}{E^0_n - E^0_k}, \nn \\
    &\alpha_{kn} \equiv \sumal{l \neq n}{}{\left(\frac{\bra{\psi^0_k} V \ket{\psi^0_l} \bra{\psi^0_l} V \ket{\psi^0_n}}{(E^0_n - E^0_l)(E^0_n - E^0_k)}\right)} - \frac{\bra{\psi^0_n} V \ket{\psi^0_n} \bra{\psi^0_k}V\ket{\psi^0_n}}{(E^0_n - E^0_k)^2}, \nn \\
\label{eq:ratios}
\end{eqnarray}
where $V$ is the interaction, $E^0_n$ and $E^0_k$ are the many-body energies of $\ket{\psi^0_n}$ and $\ket{\psi^0_k}$ respectively.
Note that in the cases we are working with, the Hamiltonian $H_0$ and interaction $V$ are time-reversal symmetric, and thus $g_{kn}$ and $\alpha_{kn}$ are real numbers, which simplifies the following analysis.
Using Eq.~(\ref{wfperturbed}) we expand the density matrix of $\ket{\psi_n}$ as
\onecolumngrid
\begin{eqnarray}
    &\ket{\psi_n}\bra{\psi_n} =\left(1 - \frac{\lambda^2}{2} \sumal{k \neq n}{}{g_{kn}^2}\right)^2 \ket{\psi^0_n}\bra{\psi^0_n} + \lambda^2\sumal{k \neq n}{}{\left[g_{kn}^2 \ket{\psi^0_k}\bra{\psi^0_k}\right]} + \frac{\lambda^2}{2} \sumal{k \neq l \neq n}{}{\left[g_{kn} g_{ln} \left(\ket{\psi^0_k}\bra{\psi^0_l} + \ket{\psi^0_l}\bra{\psi_{k}}\right)\right]} \nn \\
    &+ \left(1 - \frac{\lambda^2}{2} \sumal{k \neq n}{}{g_{kn}^2}\right)\lambda \sumal{k\neq n}{}{\left[g_{kn} \left(\ket{\psi^0_k}\bra{\psi^0_n} + \ket{\psi^0_n}\bra{\psi^0_k}\right)\right]} +\lambda^2 \sumal{k \neq n}{}{\alpha_{kn} \left(\ket{\psi^0_k}\bra{\psi^0_n} + \ket{\psi^0_n}\bra{\psi^0_k}\right)}
\label{eq:rhoexpand}
\end{eqnarray}
Defining the matrices
\begin{eqnarray}
    &\rhored \equiv \textrm{Tr}_B\left(\ket{\psi_n}\bra{\psi_n}\right),\;\;\; \rho^0_{ll} \equiv \textrm{Tr}_B\left(\ket{\psi^0_l}\bra{\psi^0_l}\right) \nn \\
    &\rho^0_{lm} \equiv \frac{1}{2}\textrm{Tr}_B\left(\ket{\psi^0_l}\bra{\psi^0_m} + \ket{\psi^0_m}\bra{\psi^0_l}\right),
\label{eq:defn}
\end{eqnarray}
where $\textrm{Tr}_B$ represents the trace over a subsystem, using Eq.~(\ref{eq:rhoexpand}) we obtain 
\begin{equation}
    \rhored = (1 - \lambda^2\sumal{k\neq n}{}{g_{kn}^2})\rho_{nn} + \lambda^2 \sumal{k \neq n}{}{\left[g_{kn}^2 \rho_{kk} + 2\alpha_{kn} \rho_{kn}\right]} + \lambda^2 \sumal{k \neq k' \neq n}{}{\left[g_{kn} g_{k'n} \rho_{k k'}\right]}  + 2 \lambda \sumal{k \neq n}{}{\left[g_{kn} \rho_{kn}\right]} + \mo(\lambda^3),
\end{equation}
and consequently
\begin{eqnarray}
    &\rhored^2 = \left[\left(1 - \lambda^2 \sumal{k \neq n}{}{g_{kn}^2}\right) \rho_{nn}\right]^2 + 4 \lambda \sumal{k \neq n}{}{g_{kn} \rho_{kn} \rho_{nn}} + 2 \lambda^2 \left(\sumal{k \neq n}{}{\left[g_{kn}^2 \rho_{kk} \rho_{nn} + 2\alpha_{kn} \rho_{kn}\rho_{nn}\right]} + \sumal{k \neq k' \neq n}{}{g_{kn} g_{k'n} \rho_{kk'} \rho_{nn}}\right) \nn \\
    &+ 4 \lambda^2 \left[\sumal{k \neq n}{}{g_{kn} \rho_{kn}}\right]^2 + \mathcal{O}\left(\lambda^2\right).
\end{eqnarray}
Restricting to $\mathcal{O}\left(\lambda^2\right)$, taking a trace, and rearranging terms, we obtain
\begin{eqnarray}
    &\textrm{Tr}\left(\rhored^2\right) = \textrm{Tr}(\rho_{nn}^2) + 4 \lambda \sumal{k \neq n}{}{\left[g_{kn} \textrm{Tr}\left(\rho_{kn} \rho_{nn}\right)\right]}+ 2\lambda^2 \left( \sumal{k \neq k' \neq n}{}{\left[ g_{kn} g_{k'n} \textrm{Tr}\ \left(\rho_{k k'} \rho_{nn} + 2 \rho_{kn} \rho_{k' n}\right)\right]}\right. \nn \\
    &\left. +  \sumal{k \neq n}{}{\left[g_{kn}^2 \left(\textrm{Tr}\left(\rho_{kk}\rho_{nn} + 2\rho_{kn}^2 - \rho_{nn}^2\right)\right) + 2\alpha_{kn} \rho_{kn} \rho_{nn} \right]}\right) + \mo\left(\lambda^3\right).\nn \\
\end{eqnarray}
Thus, using Eq.~(\ref{Sdefn}) an expression for $S(\lambda, L)$ that reads  
\begin{eqnarray}
    &S(\lambda, L) = -\log \textrm{Tr}\left(\rho^2_{nn}\right) - 4 \lambda \sumal{k}{}{\left[g_{kn} \frac{\textrm{Tr}\left(\rho_{kn}\rho_{nn}\right)}{\textrm{Tr}(\rho_{nn}^2)}\right]} + 8 \lambda^2 \left(\sumal{k \neq n}{}{\left[g_{kn} \frac{\textrm{Tr}\left(\rho_{kn}\rho_{nn}\right)}{\textrm{Tr}(\rho_{nn}^2)}\right]}\right)^2 - 2\lambda^2\left(\sumal{k \neq n}{}{\left[g_{kn}^2 \frac{\textrm{Tr}\left(\rho_{kk}\rho_{nn} + 2\rho_{kn}^2 - \rho_{nn}^2\right)}{\textrm{Tr}\left(\rho_{nn}^2\right)}\right]}\right. \nn \\
    & \left.+ 2\sumal{k \neq n}{}{\left[\alpha_{kn}\frac{\textrm{Tr}\left(\rho_{kn}\rho_{nn}\right)}{\textrm{Tr}\left(\rho_{nn}^2\right)}\right]} + \sumal{k \neq l \neq n}{}{\left[ g_{kn} g_{ln} \frac{\textrm{Tr}\ \left(\rho_{k l} \rho_{nn} + 2\rho_{kn} \rho_{l n}\right)}{\textrm{Tr}\left(\rho_{nn}^2\right)}\right]}\right)  \nn \\
    &\equiv S_0\left(L\right) + \lambda S_1\left(L\right) + \frac{\lambda^2}{2} S_2\left(L\right) + \mo\left(\lambda^3\right).
\end{eqnarray}
Consequently,
\begin{eqnarray}
    &S_1(L) = -4\sumal{k \neq n}{}{\left[g_{kn} \frac{\textrm{Tr}\left(\rho_{kn}\rho_{nn}\right)}{\textrm{Tr}(\rho_{nn}^2)}\right]}, \label{eq:S1expand} \\
    &S_2(L) =  -4\sumal{k}{}{\left[g_{kn}^2 \frac{\textrm{Tr}\left(\rho_{kk}\rho_{nn} + 2\rho_{kn}^2 - \rho_{nn}^2\right)}{\textrm{Tr}\left(\rho_{nn}^2\right)}\right]} + 16\left(\sumal{k \neq n}{}{\left[g_{kn} \frac{\textrm{Tr}\left(\rho_{kn}\rho_{nn}\right)}{\textrm{Tr}(\rho_{nn}^2)}\right]}\right)^2 - 4 \sumal{k \neq l \neq n}{}{\left[ g_{kn} g_{ln} \frac{\textrm{Tr}\left(\rho_{kl} \rho_{nn} + 2\rho_{k n} \rho_{ln}\right)}{\textrm{Tr}\left(\rho_{nn}^2\right)}\right]} \nn \\
    &+ 4\sumal{k \neq n}{}{\left[\alpha_{kn}\frac{\textrm{Tr}\left(\rho_{kn}\rho_{nn}\right)}{\textrm{Tr}\left(\rho_{nn}^2\right)}\right]}
    \label{eq:S2expand}
\end{eqnarray}
$S_2(L)$ in Eq.~(\ref{eq:S2expand}) can further be written as
\begin{eqnarray}
    &S_2(L) = \sumal{k \neq n}{}{g_{kn}^2\left( 4 - \frac{4\  \textrm{Tr}\left(\rho_{kk}\rho_{nn}\right)}{\textrm{Tr}\left(\rho_{nn}^2\right)} - \frac{8\  \textrm{Tr}\left(\rho_{kn}^2\right)}{\textrm{Tr}\left(\rho_{nn}^2\right)} + \frac{16\  \textrm{Tr}\left(\rho_{kn} \rho_{nn}\right)^2}{\textrm{Tr}\left(\rho_{nn}^2\right)^2}\right)}  \nn \\
    & + \sumal{k \neq l \neq n}{}{ g_{kn} g_{ln} \left( \frac{16\ \textrm{Tr}\left(\rho_{kn} \rho_{nn}\right)\textrm{Tr}\left(\rho_{ln} \rho_{nn}\right)}{\textrm{Tr}\left(\rho_{nn}^2\right)^2} - \frac{4\ \textrm{Tr}\left(\rho_{kl} \rho_{nn}\right)}{\textrm{Tr}\left(\rho_{nn}^2\right)} + \frac{8\ \textrm{Tr}\left(\rho_{k n} \rho_{l n}\right)}{\textrm{Tr}\left(\rho_{nn}^2\right)}\right)} + 4\sumal{k \neq n}{}\alpha_{kn}\frac{\textrm{Tr}\left(\rho_{kn}\rho_{nn}\right)}{\textrm{Tr}\left(\rho_{nn}^2\right)}\nn \\
    &\equiv \sumal{k \neq n}{}{c_{kn} g_{kn}^2} + \sumal{k \neq l \neq n}{}{d_{k l n} g_{kn} g_{ln}} + \sumal{k \neq n}{}{e_{kn} \alpha_{kn}},
\label{eq:S2final}
\end{eqnarray}
where $c_{kn}$, $d_{kln}$, and $e_{kn}$, where they are all bounded $\mathcal{O}\left(1\right)$ quantities defined as
\begin{eqnarray}
    &c_{kn} \equiv 4\left(1 - \frac{\ \textrm{Tr}\left(\rho_{kk}\rho_{nn}\right)}{\textrm{Tr}\left(\rho_{nn}^2\right)} - \frac{2\ \textrm{Tr}\left(\rho_{kn}^2\right)}{\textrm{Tr}\left(\rho_{nn}^2\right)} + \frac{4\  \textrm{Tr}\left(\rho_{kn} \rho_{nn}\right)^2}{\textrm{Tr}\left(\rho_{nn}^2\right)^2}\right) \label{eq:ckndefn} \\
    &d_{kln} = 4\left( \frac{4\ \textrm{Tr}\left(\rho_{kn} \rho_{nn}\right)\textrm{Tr}\left(\rho_{ln} \rho_{nn}\right)}{\textrm{Tr}\left(\rho_{nn}^2\right)^2} - \frac{\textrm{Tr}\left(\rho_{kl} \rho_{nn}\right)}{\textrm{Tr}\left(\rho_{nn}^2\right)} + \frac{2\ \textrm{Tr}\left(\rho_{k n} \rho_{l n}\right)}{\textrm{Tr}\left(\rho_{nn}^2\right)}\right) \label{eq:dklndefn} \\
    &e_{kn} = \frac{\textrm{Tr}\left(\rho_{kn}\rho_{nn}\right)}{\textrm{Tr}\left(\rho_{nn}^2\right)} \label{eq:ekndefn}.
\end{eqnarray}

\twocolumngrid

\section{Computation of matrix elements and energy gaps in the non-interacting model}\label{sec:matrixelements}
In this section we exemplify the computation of matrix elements of the interaction and energy differences between the eigenstates of the non-interacting Hamiltonian of Eq.~(\ref{eq:disorderedtfim}) (Eq.~(\ref{free}) in the Majorana fermion language).
We now recall the construction of many-body eigenstates of the non-interacting Hamiltonian.
$H$ in Eq.~(\ref{free}) can be written as a $2L \times 2L$ Hermitian matrix in the basis of the Majorana fermions as 
\begin{equation}
    H = \chi^\dagger M \chi,
\end{equation}
where $\chi$ is $(\chi_1\  \chi_2\ \dots\  \chi_{2L})^T$, a vector of Majorana fermions.
$M$ is Hermitian as well as anti-symmetric, thus its eigenvalues occur in pairs of real $(+\mathcal{E}, -\mathcal{E})$, where we assume $\mathcal{E} \geq 0$.
If the eigendecomposition of $M$ reads
\begin{equation}
M = Q \Lambda Q^\dagger,
\end{equation}
where $\Lambda$ is a diagonal, we obtain 
\begin{equation}
    H = \chi^\dagger Q \Lambda Q^\dagger \chi.
\label{eq:Heig}
\end{equation}
The creation and annihilation operators of the single-particle eigenstates of $H$ $\{b_n\}$ and $\{b_n^\dagger\}$ are then encoded in the vectors $b = (b_1\ b_2\ \cdots\ b_L\ b_1^\dagger\ b_2^\dagger\ \cdots\ b_L^\dagger)^T$ and $b^\dagger$, where
\begin{equation}
b = Q^\dagger \chi, \;\;\; b^\dagger = \chi^\dagger Q.
\label{eq:bbdag}
\end{equation}
Thus the eigenstates corresponding to single-particle energies $+\mathcal{E}_\alpha$ and $-\mathcal{E}_\alpha$ are $b_\alpha^\dagger\ket{0}$ and  $b_\alpha\ket{0}$, where $\ket{0}$ is the Fock vacuum that satisfies $c_j \ket{0}\;\;\forall j$. 
Using Eq.~(\ref{eq:bbdag}), with an appropriate labelling of the indices of $Q$, the Majorana fermions can be written in terms of $b_\alpha$'s and $b_\alpha^\dagger$'s as 
\begin{equation}
    \chi_i = \sum_{\alpha=1}^L{\left(Q_{i,\alpha} b^\dagger_\alpha + Q_{i,-\alpha} b_{\alpha}\right)} \equiv \sum_{\alpha \neq 0}{}{Q_{i,\alpha}b^t_\alpha},
\label{majoranas}
\end{equation}
where for convenience we introduced the notation
\begin{equation}
    b^t_\alpha \equiv \twopartdef{b_{|\alpha|}^\dagger}{\alpha > 0}{b_{|\alpha|}}{ \alpha < 0},
\label{eq:btdefn}
\end{equation}
In Eq.~(\ref{majoranas}), since $\chi_i$'s are real, the $Q_{i,\alpha}$'s satisfy
\begin{equation}
    Q_{i,-\alpha}^\ast = Q_{i,\alpha}.   
\label{eq:Qiprop}
\end{equation}
The interaction of Eq.~(\ref{interaction}) at the critical point can then be written in terms of the $b^t_\alpha$'s as 
\begin{eqnarray}
    V &=& C \sumal{i,\alpha,\beta,\gamma,\delta}{}{Q_{i,\alpha}Q_{i+1,\beta}Q_{i+2,\gamma}Q_{i+3,\delta} b^t_\alpha b^t_\beta b^t_\gamma b^t_\delta} \nn \\
    &\equiv& \sumal{\alpha,\beta,\gamma,\delta}{}{V_{\left(\alpha, \beta, \gamma, \delta\right)}b^t_\alpha b^t_\beta b^t_\gamma b^t_\delta}, 
\label{int}
\end{eqnarray}
where we have defined
\begin{equation}
    V_{\left(\alpha, \beta, \gamma, \delta\right)} \equiv C \sumal{i}{}{Q_{i,\alpha}Q_{i+1,\beta}Q_{i+2,\gamma}Q_{i+3,\delta}}.
\end{equation}
A similar expression can be obtained for the interaction away from the critical point. 
Using Eq.~(\ref{int}), we want to obtain matrix elements between many-body eigenstates $\ket{\psi^0_n}$ and $\ket{\psi^0_k}$. That is, we want to compute
\begin{equation}
    V_{kn} \equiv \bra{\psi^0_k}V\ket{\psi^0_n} = \sumal{\alpha,\beta,\gamma,\delta}{}{V_{\left(\alpha,\beta,\gamma,\delta\right)} \bra{\psi^0_k} b^t_\alpha b^t_\beta b^t_\gamma b^t_\delta \ket{\psi^0_n}}.
\label{eq:Vmatrixel}
\end{equation}
In Eq.~(\ref{eq:Vmatrixel}), it is clear that all the matrix elements $\bra{\psi^0_k} b^t_\alpha b^t_\beta b^t_\gamma b^t_\delta \ket{\psi^0_n}$ vanish unless $\ket{\psi^0_n}$ and $\ket{\psi^0_k}$ differ in the occupation of four or two of the single-particle orbitals, the computation of which we illustrate separately. 
We first introduce the notations and conventions used in the following subsections.
For any tuple $A = (x_1, x_2, \cdots, x_n)$, we introduce
\begin{equation}
-A \equiv (-x_n, -x_{n-1}, \cdots, -x_1). 
\end{equation}
We further introduce products 
\begin{eqnarray}
    &&\prodal{x \in A}{\rightarrow}{f(x)} = f(x_1) f(x_2) \cdots f(x_n),\nn \\
    &&\prodal{x \in A}{\leftarrow}{f(x)} = f(x_n) f(x_{n-1})\cdots f(x_1).
\end{eqnarray}
We also introduce a tuple concatenation operator $\circ$ that acts on tuples $A = \left(x_1, x_2, \cdots, x_{n_A}\right)$ and $B = \left(y_1, y_2, \cdots, y_{n_B}\right)$ as
\begin{equation}
    C = A \circ B = \left(x_1, x_2, \cdots, x_{n_A}, y_1, y_2, \cdots, y_{n_B}\right). 
\end{equation}
We will also be using the usual set operation $/$ (difference) for tuples instead.
Furthermore, for any $n$-tuple $A$, for any permutation $\sigma$ that belongs to the permutation group $S_{n}$, we denote the corresponding permutation of $A$ as $\sigma(A)$. 
Finally, we define tuples $\{\Lambda_n\}$ with elements ordered in ascending order as the tuple containing the indices of the single-particle orbitals that are occupied in the many-body states $\{\ket{\psi^0_n}\}$. 
Consequently, the expression for the many-body state $\ket{\psi^0_n}$ reads
\begin{equation}
    \ket{\psi^0_n} = \prodal{\alpha \in \Lambda_n}{\rightarrow}{b^\dagger_\alpha}\ket{\theta},
\label{eq:canonical}
\end{equation}
where $\ket{\theta}$ is the Bogoliubov vacuum defined by
\begin{equation}
    b_\alpha \ket{\theta} = 0\;\;\; \forall \alpha.
\label{eq:bogvacuum}
\end{equation}
\onecolumngrid
\subsection{Four-orbital processes}
We first consider the matrix element of the interaction between two many-body eigenstates $\ket{\psi^0_n}$ and $\ket{\psi^0_k}$ which differ in the occupation of four of the single-particle orbitals, say the orbitals $\mo = \left(p,q,r,s\right)$ where $0 < p < q < r < s$. 
Here, the matrix element can be non-zero only if $\ket{\psi^0_n}$ and $\ket{\psi^0_k}$ are of the forms 
\begin{equation}
    \ket{\psi^0_n} = \prodal{\alpha \in \Lambda_n}{\rightarrow}{b^\dagger_{\alpha}}\ket{\theta} = \underbrace{(-1)^{\sumal{\alpha \in \mo_n}{}{\sumal{\beta = 1}{\alpha - 1}{n_\beta}}}}_{\eta_n}\prodal{\alpha \in \mo_n}{\rightarrow}{b^\dagger_\alpha}\ket{\psi},\;\;\;\ket{\psi^0_k} = \prodal{\alpha \in \Lambda_k}{\rightarrow}{b^\dagger_{\alpha}}\ket{\theta} = \underbrace{(-1)^{\sumal{\alpha \in \mo_k}{}{\sumal{\beta = 1}{\alpha - 1}{n_\beta}}}}_{\eta_k}\prodal{\alpha \in \mo_k}{\rightarrow}{b^\dagger_{\alpha}}\ket{\psi},
\label{eq:psikpsindefn}
\end{equation}
where $\mo_n, \mo_k \subseteq \mo$ are disjoint tuples such that $\mo_n = \mo/\mo_k$, and  $\ket{\psi}$ is a many-body eigenstate in which the orbitals $p$, $q$, $r$ and $s$ are unoccupied, i.e.
\begin{equation}
    \ket{\psi} = \prodal{\alpha \in \Lambda_n/\mo_n}{\rightarrow}{b^\dagger_\alpha}\ket{\theta},
\label{eq:psieq}
\end{equation}
and $n_\beta$ is the occupation number of the single-particle orbital $\beta$ in $\ket{\psi}$.
The matrix element of Eq.~(\ref{eq:Vmatrixel}) then reads
\begin{eqnarray}
    V_{kn} &=& \eta_k \eta_n \sumal{\sigma \in S_4}{}{\left(V_{\sigma\left(\mo_k \circ -\mo_n \right)} \bra{\psi} \prodal{\alpha \in \mo_k}{\leftarrow}{b_\alpha} \prodal{\alpha \in \sigma\left(\mo_k \circ -\mo_n \right)}{\rightarrow}{b^t_\alpha} \prodal{\alpha \in \mo_n}{\rightarrow}{b^\dagger_\alpha} \ket{\psi}\right)} \nn\\
    &=& (-1)^{\sumal{\alpha \in \mo_k}{}{\sumal{\beta = 1}{\alpha - 1}{n_\beta}} + \sumal{\alpha \in \mo_n}{}{\sumal{\beta = 1}{\alpha - 1}{n_\beta}}}  \sumal{\sigma \in S_4}{}{\left(\textrm{sgn}\left(\sigma\right)V_{\sigma\left(\mo_k \circ -\mo_n \right)}\right)  \bra{\psi}\prodal{\alpha \in \mo_k}{\leftarrow}{b_\alpha} \prodal{\alpha \in \mo_k}{\rightarrow}{b^\dagger_\alpha} \prodal{\alpha \in \mo_n}{\leftarrow}{b_\alpha} \prodal{\alpha \in \mo_n}{\rightarrow}{b^\dagger_\alpha} \ket{\psi}} \nn \\
    &=& (-1)^{\sumal{\alpha \in \mo}{}{\sumal{\beta = 1}{\alpha - 1}{n_\beta}}}  \sumal{\sigma \in S_4}{}{\left(\textrm{sgn}\left(\sigma\right)V_{\sigma(\mo_k \circ -\mo_n)}\right)} = (-1)^{\sumal{\beta = p}{q - 1}{n_\beta} + \sumal{\beta = r}{s - 1}{n_\beta}}  \sumal{\sigma \in S_4}{}{\left(\textrm{sgn}\left(\sigma\right)V_{\sigma(\mo_k \circ -\mo_n)}\right)}.
\label{eq:matrixfin4}
\end{eqnarray}
The energy differences between $E^0_n$ and $E^0_k$ then read
\begin{equation}
    E_{kn} \equiv E^0_n - E^0_k = \sumal{\alpha \in \mo_n}{}{\mathcal{E}_\alpha} - \sumal{\alpha \in \mo_k}{}{\mathcal{E}_\alpha}.
\label{eq:gapfin4}
\end{equation}
An important observation is that the magnitudes of $V_{kn}$ and $E_{kn}$ in Eqs.~(\ref{eq:matrixfin4}) and (\ref{eq:gapfin4}) do not depend on the occupations of any of the single-particle orbitals apart from the ones involved.  
\subsection{Two-orbital processes}
We now discuss the case where $\ket{\psi^0_n}$ and $\ket{\psi^0_k}$ differ in the occupation of two single-particle orbitals, $\mo = \left(p, q\right)$ where $0 < p < q$.
Here too, the matrix element of the interaction can be non-zero only if $\ket{\psi^0_k}$ and $\ket{\psi^0_n}$ are of the forms shown in Eq.~(\ref{eq:psikpsindefn}), where $\mo = \left(p, q\right)$.
The two-orbital matrix element of the interaction operator reads 
\begin{eqnarray}
    V_{kn} &=& \eta_k \eta_n  \sumal{\gamma}{}{\sumal{\sigma \in S_4}{}{\left(V_{\sigma\left(\mo_k \circ \left(\gamma, -\gamma\right) \circ -\mo_n \right)} \bra{\psi} \prodal{\alpha \in \mo_k}{\leftarrow}{b_\alpha} \prodal{\alpha \in \sigma\left(\mo_k \circ \left(\gamma, -\gamma\right) \circ -\mo_n \right)}{\rightarrow}{b^t_\alpha} \prodal{\alpha \in \mo_n}{\rightarrow}{b^\dagger_\alpha} \ket{\psi}\right)}}\nn \\
    &=& \eta_k \eta_n \left[ \sumal{\gamma}{}{\sumal{\sigma \in S_4}{}{\left(\textrm{sgn}\left(\sigma \right)V_{\sigma\left(\mo_k \circ \left(\gamma, -\gamma\right) \circ -\mo_n\right)}\right)}}\right. \bra{\psi}\prodal{\alpha \in \mo_k}{\leftarrow}{b_\alpha} \prodal{\alpha \in \mo_k}{\rightarrow}{b^\dagger_\alpha}\; \left(b^\dagger_\gamma b_\gamma\right) \prodal{\alpha \in \mo_n}{\leftarrow}{b_\alpha} \prodal{\alpha \in \mo_n}{\rightarrow}{b^\dagger_\alpha} \ket{\psi} \Theta\left(\sigma\left(\mo_k \circ \left(\gamma, -\gamma\right) \circ -\mo_n\right), \gamma, -\gamma\right) \nn \\
    &&+\sumal{\gamma}{}{\sumal{\sigma \in S_4}{}{\left(\textrm{sgn}\left(\sigma\right)V_{\sigma\left(\mo_k \circ \left(-\gamma, \gamma\right) \circ -\mo_n\right)}\right)}} \left. \bra{\psi}\prodal{\alpha \in \mo_k}{\leftarrow}{b_\alpha} \prodal{\alpha \in \mo_k}{\rightarrow}{b^\dagger_\alpha}\; \left(b_\gamma b^\dagger_\gamma\right) \prodal{\alpha \in \mo_n}{\leftarrow}{b_\alpha} \prodal{\alpha \in \mo_n}{\rightarrow}{b^\dagger_\alpha} \ket{\psi} \Theta\left(\sigma\left(\mo_k \circ \left(-\gamma, \gamma\right) \circ -\mo_n\right), -\gamma, \gamma\right)\right], \nn \\
\label{eq:matrixmid2}
\end{eqnarray}
where we have defined a $\Theta$ symbol for a tuple $A$ and elements $a$, $b$ of the tuple as
\begin{equation}
    \Theta\left(A, a, b\right) = \twopartdef{1}{\textrm{$a$ appears to the left of $b$ in $A$}}{0}{\textrm{$a$ appears to the right of $b$ in $A$}}.
\end{equation}
Thus, the matrix element of Eq.~(\ref{eq:matrixmid2}) can be written as
\begin{eqnarray}
    V_{kn} &=& (-1)^{\sumal{\alpha \in \mo_k}{}{\sumal{\beta = 1}{\alpha - 1}{n_\beta}} + \sumal{\alpha \in \mo_n}{}{\sumal{\beta = 1}{\alpha - 1}{n_\beta}}} \left[\sumal{\gamma}{}{\sumal{\sigma \in S_4}{}{\left(\textrm{sgn}\left(\sigma\right)V_{\sigma(\mo_k \circ \left(\gamma, -\gamma\right) \circ -\mo_n)}\delta_{n_\gamma, 0} \Theta\left(\sigma\left(\mo_k \circ \left(\gamma, -\gamma\right) \circ -\mo_n\right), \gamma, -\gamma\right) \right)}}\right. \nn \\
    &&\left.+ \sumal{\gamma}{}{\sumal{\sigma \in S_4}{}{\left(\textrm{sgn}\left(\sigma\right)V_{\sigma(\mo_k \circ \left(-\gamma, \gamma\right) \circ -\mo_n)}\delta_{n_\gamma, 1}\Theta\left(\sigma\left(\mo_k \circ \left(-\gamma, \gamma\right) \circ -\mo_n\right), -\gamma, \gamma\right)\right)}}\right] \nn \\
    &=& (-1)^{\sumal{\alpha \in \mo}{}{\sumal{\beta = 1}{\alpha - 1}{n_\beta}}}  \left[\sumal{\gamma}{}{\sumal{\sigma \in S_4}{}{\left(\textrm{sgn}\left(\sigma\right)V_{\sigma(\mo_k \circ \left(\gamma, -\gamma\right) \circ -\mo_n)} \delta_{n_\gamma, 0} \Theta\left(\sigma\left(\mo_k \circ \left(\gamma, -\gamma\right) \circ -\mo_n\right), \gamma, -\gamma\right)\right)} } \right.\nn \\
    &&\left.- \sumal{\gamma}{}{\sumal{\sigma \in S_4}{}{\left(\textrm{sgn}\left(\sigma\right)V_{\sigma(\mo_k \circ \left(\gamma, -\gamma\right) \circ -\mo_n)}\delta_{n_\gamma, 1}\Theta\left(\sigma\left(\mo_k \circ \left(\gamma, -\gamma\right) \circ -\mo_n\right), -\gamma, \gamma\right)\right)}}\right] \nn \\
    &=& (-1)^{\sumal{\beta = p}{q - 1}{n_\beta}} \sumal{\gamma}{}{(-1)^{n_\gamma}\sumal{\sigma \in S_4}{}{\left(\textrm{sgn}\left(\sigma\right)V_{\sigma(\mo_k \circ \left(\gamma, -\gamma\right) \circ -\mo_n)}\Theta\left(\sigma\left(\mo_k \circ \left(\gamma, -\gamma\right) \circ -\mo_n\right), (-1)^{n_\gamma}\gamma, (-1)^{n_\gamma + 1}\gamma\right)\right)}},\nn \\
\label{eq:matrixfin2}
\end{eqnarray}
where we have used the facts that
\begin{equation}
    \bra{\psi}\prodal{\alpha \in \mo_k}{\leftarrow}{b_\alpha} \prodal{\alpha \in \mo_k}{\rightarrow}{b^\dagger_\alpha}\; \left(b^\dagger_\gamma b_\gamma\right) \prodal{\alpha \in \mo_n}{\leftarrow}{b_\alpha} \prodal{\alpha \in \mo_n}{\rightarrow}{b^\dagger_\alpha} \ket{\psi} = \delta_{n_\gamma, 0},\;\;\;\bra{\psi}\prodal{\alpha \in \mo_k}{\leftarrow}{b_\alpha} \prodal{\alpha \in \mo_k}{\rightarrow}{b^\dagger_\alpha}\; \left(b_\gamma b^\dagger_\gamma\right) \prodal{\alpha \in \mo_n}{\leftarrow}{b_\alpha} \prodal{\alpha \in \mo_n}{\rightarrow}{b^\dagger_\alpha} \ket{\psi} =  \delta_{n_\gamma, 1}.
\end{equation}
Meanwhile, the energy difference between $E^0_n$ and $E^0_k$ reads
\begin{equation}
    E_{kn} \equiv E^0_n - E^0_k = \sumal{\alpha \in \mo_n}{}{\mathcal{E}_\alpha} - \sumal{\alpha \in \mo_k}{}{\mathcal{E}_\alpha}.
\label{eq:gapfin2}
\end{equation}
Unlike four-orbital processes, the magnitude of $V_{kn}$ in Eq.~(\ref{eq:matrixfin2}) does depend on the occupations of the orbitals other than the ones directly involved in the process. 
\twocolumngrid

\section{Review of the non-interacting model}\label{app:nonint}
\begin{figure*}[ht!]
\centering
\includegraphics[scale = 1]{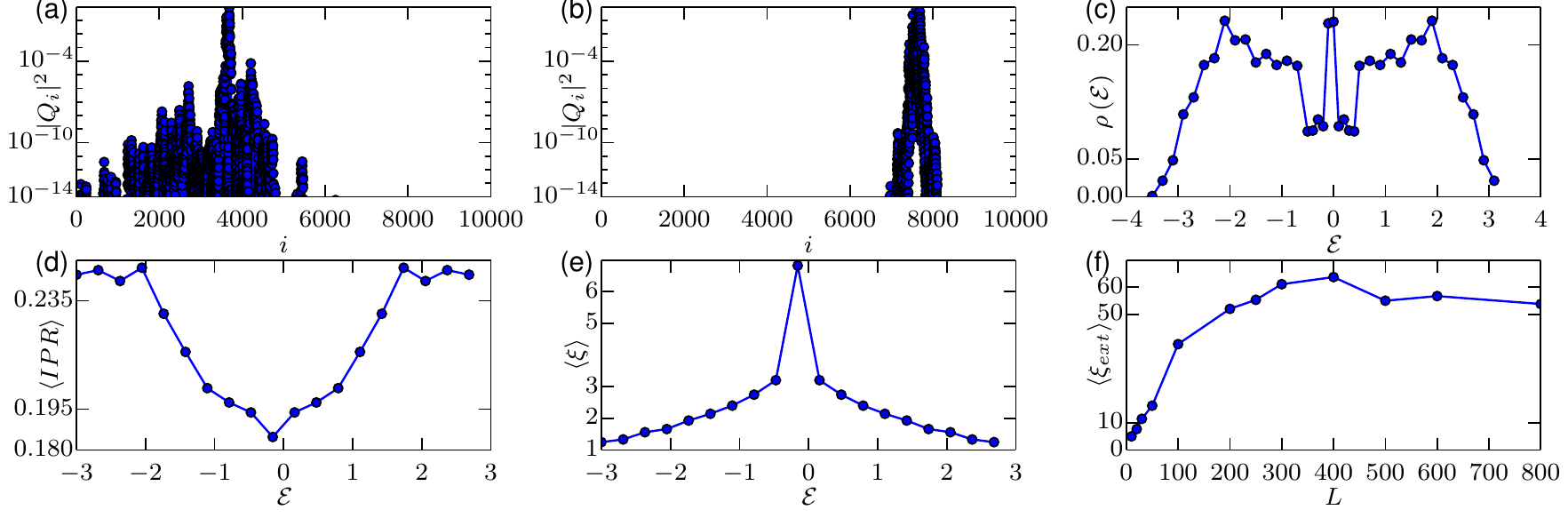}
\caption{Various properties of the non-interacting model at the critical point $\Delta_{Jh} = 0$ or $\log{\Jm} = \log{\hm}$ (a-b) Spatial profile of a typical single-particle eigenstate with energy (a) $\mE \approx 0$ (b) $\mE \sim \mathcal{O}(1)$ for $L = 5000$. Figures show the weights $|Q_i|^2$ of the eigenstates on the $i$'th Majorana fermion $\chi_i$. (c) Single-particle density of states (d) Typical Inverse Participation Ratio (IPR) of the single-particle eigenstates. A lower IPR indicates lesser localization.  (e) Typical second moment of the single-particle eigenstates. (f) Growth of the typical second moment of the single-particle eigenstates closest to $\mE = 0$ with system size $L$.}
\label{fig:SPdata}
\end{figure*}
In this appendix, we review the properties of the non-interacting ($\lambda = 0$) limit of the Hamiltonian Eq.~(\ref{hamil}).
In this limit, the Hamiltonian is the well-known disordered transverse field Ising model of Eq.~(\ref{eq:disorderedtfim}) which has been widely studied in literature in various contexts~\cite{mccoy1968theory, fisher1995critical, fisher1998distributions, young1996numerical, igloi1998random, motrunich2000infinite, mckenzie1996exact, degottardi2013majorana}. 
For concreteness, we assume uniform distributions in the couplings $\{J_i\} \in [0, \Jm]$ and fields $\{h_i\} \in [0, \hm]$ in Eq.~(\ref{eq:disorderedtfim}) and open boundary conditions. 
Since $H_0$ is statistically Kramers-Wannier self-dual, its eigenstates undergo a phase transition between a ``spin-glass" phase to the paramagnetic phase at $\log\Jm = \log\hm$.
As shown in Eq.~(\ref{free}) of App.~\ref{app:JWtransformation}, the Hamiltonian of Eq.~(\ref{eq:disorderedtfim}) can be written as a quadratic Hamiltonian of $2L$ Majorana fermions.
The properties of $H_0$ can thus be understood using the single particle eigenstates of Eq.~(\ref{free}).
The model of Eq.~(\ref{eq:disorderedtfim}) has been solved using the Strong Disorder Renormalization Group (SDRG), a real-space renormalization group~\cite{fisher1995critical, refael2013strong}.
This RG procedure proceeds by diagonalizing the strongest on-site/nearest-neighbor term in the Hamiltonian, projecting onto its low-energy subspace and decimating the site/bond associated with that term.
Each step of the RG moves lower in energy and changes the distributions of the couplings $\{J_i\}$ and the fields $\{h_i\}$~\cite{refael2013strong}. 
The nature of the RG fixed point is different at the critical point (where $\{h_i\}$ and $\{J_i\}$ are chosen from identical distributions) and away from the critical point~\cite{fisher1995critical}.
At the critical point, typical end-end correlation functions $\langle \sigma^x_1 \sigma^x_L \rangle_{\textrm{GS}}$ in the many-body ground state scale \textit{stretched exponentially}, and so does the typical energy gap $\Delta E_{\textrm{GS}}$ above the many-body ground state~\cite{fisher1998distributions,fisher1995critical,refael2013strong, young1996numerical}.
That is, for a system size of $L$ with open boundary conditions, 
\begin{equation}
    \langle \langle \sigma^x_1 \sigma^x_L \rangle_{\textrm{GS}}\rangle_{\textrm{typ}} \sim e^{-\sqrt{\frac{L}{\Xi}}},\; \langle \Delta E_{\textrm{GS}} \rangle_{\textrm{typ}} \sim e^{-\sqrt{\frac{L}{\Delta}}},
\label{CPscaling}
\end{equation}
where the distributions of $\Xi$ and $\Delta$ are derived in Ref.~[\onlinecite{fisher1998distributions}].
In terms of the single-particle eigenstates of Eq.~(\ref{free}), we find that a few of the single-particle eigenstates close to $\mE = 0$ are ``extended", or more precisely stretched exponentially localized at the critical point, consistent with the scaling of the typical correlations in Eq.~(\ref{CPscaling}).
That is, we find that the wavefunction of the $\alpha$-th single-particle eigenstate (one with energy $\mE_\alpha$) has the form
\begin{equation}
    |\psi_\alpha(x)| \sim \twopartdefoth{\exp\left({-\sqrt{\frac{|x - R_\alpha|}{\xe}}}\right)}{\mE_\alpha \approx 0}{\exp\left({-\frac{|x - R_\alpha|}{\xl}}\right)},
\label{eq:psiformcrit}
\end{equation}
where $R_\alpha$ is the localization center, $\xe$ and $\xl$ are the second moments of the wavefunctions.  
Examples of single-particle wavefunctions at the critical point close to and away from $\mE = 0$ are shown in Figs.~\ref{fig:SPdata}a-b.
Indeed, the typical IPR's (resp. second moments) of the single-particle orbitals appear to decrease (resp. increase), as shown in Fig.~\ref{fig:SPdata}d (resp. Fig.~\ref{fig:SPdata}e).
However, the second moments of the stretched exponentially localized orbitals saturate to a constant for large $L$.
The growth of the second moment of the single-particle orbital closest to $\mE = 0$ with system size is shown in Fig.~\ref{fig:SPdata}f.
Furthermore, as shown in Fig.~\ref{fig:SPdata}c the single-particle density of states at the critical point diverges as $\mE \rightarrow 0$, and scales as
\begin{equation}
    \rho(\mE) \sim \frac{dk}{d\mE} \sim \frac{1}{L^2}\frac{dL}{d\mE} \sim -\frac{1}{\mE\left(\log \mE\right)^3},
\label{eq:dosE0}
\end{equation}
where $k$ denotes momentum. 
These properties can also be derived using the fact that the (positive part of the) single-particle spectrum of Eq.~(\ref{free}) at the critical point is identical to that of the well-studied one-dimensional fermion random hopping model~\cite{eggarter1978singular, fisher1994random, zhou2003one, krishna2020beyond}. 
Meanwhile, away from the critical point in the Hamiltonian Eq.~(\ref{eq:disorderedtfim}), the typical correlation functions and energy gaps in the ground state \cite{young1996numerical, fisher1995critical}
\begin{equation}
    \langle \langle \sigma^x_1 \sigma^x_L \rangle_{\textrm{GS}} \rangle_{\textrm{typ}} \sim \exp\left(-\frac{L}{M}\right),\;\;\; \langle \Delta E_{\textrm{GS}}\rangle_{\textrm{typ}} \sim \frac{1}{L^{\delta}}.
\label{ACPscaling}
\end{equation}
Furthermore, the single-particle spectrum away from the critical point shows a uniform density of states, and the single-particle eigenstates are all exponentially localized, with the form
\begin{equation}
    |\psi_\alpha(x)| \sim \exp\left(-\frac{|x - R_\alpha|}{\xl}\right),
\label{eq:psiformaway}
\end{equation}
where $R_\alpha$ is the localization center and $\xl$ is the second moment of the wavefunction.
%
%
\section{Nandkishore-Potter delocalization mechanism}\label{app:NP}
In this appendix, we briefly comment on a delocalization mechanism due to resonances mediated by extended states in the single-particle spectrum, as exemplified by Nandkishore and Potter (NP) in Ref.~[\onlinecite{nandkishore2014marginal}].
NP considered non-interacting fermion models where the single-particle energy eigenstates are exponentially localized with a localization length (i.e. the second moment) that scales with the single-particle energy $\mE$ as $\xi(\mE) \sim \mE^{-\nu}$ for some $\nu > 0$ such that it diverges as $\mE \rightarrow 0$.
In summary, they show that single-particle orbitals localized far away from each other hybridize at second order in perturbation theory via a hopping process mediated by the extended states at $\mE = 0$, showing that delocalization is inevitable if $\nu d > 1 + \Upsilon$, where $d$ is the dimension of the system and the single-particle density of states as $\mE \rightarrow 0$ scales as $\rho(\mE) \sim \mE^\Upsilon$. 
To apply the NP argument to the Hamiltonian Eq.~(\ref{eq:disorderedtfim}), note that the (positive part of the) single-particle spectrum of the non-interacting model Eq.~(\ref{eq:disorderedtfim}) is identical to that of a non-interacting model of spinless fermions with random hopping strengths~\cite{zhou2003one}.
This is evident when  is written in terms of Majorana fermions, shown in Eq.~(\ref{free}). 
The localization length of the single-particle eigenstates in these models thus scales as $-\log \mE$, as can be derived from Eq.~(\ref{eq:dosE0}) using the Thouless theorem that relates the single-particle density of states to the localization length in one-dimensional random hopping systems~\cite{thouless1972relation, zhou2003one}.
Thus, since as $\mE \rightarrow 0$ we have
\begin{equation}
    \rho\left(\mE\right) \sim -\frac{1}{\mE \left(\log \mE\right)^3},\;\;\xi\left(\mE\right) \sim -\log\mE
\label{eq:scalings}
\end{equation}
one might naively conclude that $\Upsilon = -1$ and $\nu = 0$ for the present model, which is marginal according to NP.
However, a closer look shows that this model is not delocalized due to NP.
We first reproduce the intuitive version of the NP condition~\cite{nandkishore2014marginal} with a slight generalization.
Using Eq.~(\ref{eq:scalings}), the number of orbitals $N(R)$ within a distance $R$ that a given localized orbital can be connected in second-order perturbation theory via the mediation of extended orbitals close to $\mE \rightarrow 0$ can be estimated as follows:~\cite{nandkishore2014marginal}
\begin{equation}
    N(R) \sim \int_{0}^{\mE_{\textrm{max}}\left(R\right)}{\mathrm{d}\mE\ \rho\left(\mE\right) \xi\left(\mE\right)}
\label{eq:numberorb}
\end{equation}
where $\mE_{\textrm{max}}\left(R\right)$ is the maximum energy for which the localization length $\xi\left(\mE\right) \geq R$, and the factor of $\xi(\mE)$ in Eq.~(\ref{eq:numberorb}) should be interpreted as the number of orbitals of energy $\mE$ that connect to a given localized orbital.
Note that $\mE_{\textrm{max}}\left(R\right) \sim \exp(-R)$ using Eq.~(\ref{eq:scalings}), 
Thus, using Eqs.~(\ref{eq:scalings}) and (\ref{eq:numberorb}), we obtain
\begin{equation}
    N(R) \sim \int_{0}^{e^{-R}}{\frac{\mathrm{d}\mE}{\mE \left(\log\mE\right)^2}} \sim \frac{1}{R}.
\end{equation}
Thus, the number of orbitals that a given orbital connects to does not increase with increasing $R$, and a delocalization at the critical point in the present model is different from the Nandkishore-Potter mechanism.  

\bibliography{ising_bib}

\end{document}